\documentclass[12pt]{article}

\usepackage{amsmath,amsfonts,enumerate,array,amssymb}
\usepackage{graphicx}
\usepackage[usenames]{color}
\usepackage[english]{babel}
\usepackage{fancybox}
\usepackage{chngpage} 
\usepackage{subcaption,cite,xcolor,lipsum}
\usepackage{tikz}

\newcommand{\m}{\mu}

\newcommand{\nn}{\nonumber\\} 		
\newcommand{\h}{{1\over2}}				

\def\h{\mathbb{h}}

\def\one{{\hbox{\kern+.5mm 1\kern-.8mm l}}}
\def\zero{{\hbox{0\kern-1.5mm 0}}}

\renewcommand{\ell}{l}

\newcommand{\ba}{\begin{align}}
\newcommand{\ea}{\begin{align}}

\newcommand{\captionfonts}{\small}

\makeatletter  
\long\def\@makecaption#1#2{%
  \vskip\abovecaptionskip
  \sbox\@tempboxa{{\captionfonts #1: #2}}%
 \ifdim \wd\@tempboxa >\hsize
    {\captionfonts #1: #2\par}
  \else
    \hbox to\hsize{\hfil\box\@tempboxa\hfil}%
  \fi
  \vskip\belowcaptionskip}
\makeatother   

\usepackage{mciteplus}

\usepackage[colorlinks=true,      linkcolor=blue,      urlcolor=blue,      filecolor=blue,      citecolor=blue,
      pdfstartview=FitH,     pdfpagemode=UseNone,      bookmarksopen=true]{hyperref}  
      
\usepackage[all]{hypcap}  
      
\begin{document}

\numberwithin{equation}{section}

\def\b{\bigskip}
\def\p{\partial}
\def\h{{1\over 2}}
\def\be{\begin{equation}}
\def\bea{\begin{eqnarray}}
\def\ee{\end{equation}}
\def\eea{\end{eqnarray}}
\def\nn{\nonumber \\}
\def\m{\medskip}
\def\r{\rightarrow}
\def\t{\tilde}

\vspace{16mm}

 \begin{center}
{\LARGE What does the information paradox say about the universe?}
\\
\vspace{18mm}
{\bf   Samir D. Mathur}

\vspace{8mm}
Department of Physics,\\ The Ohio State University,\\ Columbus,
OH 43210, USA\\ \vspace{2mm}

mathur.16@osu.edu

\vspace{8mm}
\end{center}

\vspace{4mm}

\thispagestyle{empty}
\begin{abstract}

\vspace{3mm}

The black hole information paradox is resolved in string theory by a radical change in the picture of the hole: black hole microstates are horizon sized quantum gravity objects called `fuzzballs' instead of vacuum regions with a central singularity. The requirement of causality implies that the quantum gravity wavefunctional $\Psi$ has an important component not present in the semiclassical picture: virtual fuzzballs. 
The large mass $M$ of the fuzzballs would suppress their virtual fluctuations, but this suppression is compensated  by the large number -- $Exp[S_{bek}(M)]$ -- of possible fuzzballs.  These fuzzballs are extended compression-resistant objects. The presence of these objects  in the vacuum wavefunctional  alters the physics of collapse when a horizon is about to form; this resolves the information paradox. We argue that these virtual fuzzballs also resist the curving of spacetime, and so cancel out the large cosmological constant created by the vacuum energy of local quantum fields. Assuming that the Birkoff theorem holds to leading order, we can map the  black hole information problem to a problem in cosmology. Using the virtual fuzzball component of the wavefunctional, we give a qualitative picture of the evolution of $\Psi$ which is consistent with the requirements placed by the information paradox.

\end{abstract}

\b

{\it Expanded version of the proceedings of the conference {\bf `The Physical Universe'}, Nagpur, March 2018}

\newpage

{\hypersetup{linkcolor=black}
\tableofcontents}

\section{Introduction}\label{Intro}

%
%


Classical or semiclassical gravity is an adequate approximation in most situations. Quantum gravity is expected to be relevant when the classical metric has a singularity. There are two main cases where such a singularity is of interest: in black holes, and at the big bang.

In string theory we have learnt that quantum gravity changes the entire interior of the horizon to generate a `fuzzball', and in the process  the singularity is removed \cite{mat-fuzzballs}. In this article we look at the early universe and ask what lessons we can draw for the big bang  from the fuzzball paradigm.

The plan of this article is as follows. In section \ref{sec1} we begin with a philosophical question about the universe. In section \ref{sectwo} we use the Birkoff theorem to relate the information paradox to cosmology, and thereby get a sharp puzzle. In section \ref{secc} we define two different ways of studying the universe: as a strictly infinite system, and through the  limit of a finite but large ball. In section \ref{secfour} we collect lessons from the study of the information paradox: the small corrections theorem, the constraint of causality, and the conjecture of fuzzball complementarity. In section \ref{secp} we give our proposal for the behavior of virtual fuzzballs, in the form of properties (F1)-(F5). In section \ref{secsix} we conjecture the nature of the quantum gravity wavefunctional and its evolution in the case where the universe is modeled as  a large but finite ball. Section \ref{secseven} contains our conjecture on how the high curvature implied by the large $\Lambda$ from local quantum fields is reduced to a small curvature due to the presence of virtual fuzzballs in the wavefunctional $\Psi$. We also see that our proposal about virtual fuzzballs resolves the `bags of gold' difficulty with the black hole spacetime. In section \ref{seccone} we conjecture the nature of the quantum gravity wavefunctional and its evolution in the case where the universe is infinite. Section \ref{secsummary} is a summary of our picture.

\section{A philosophical question}\label{sec1}

Consider for simplicity a flat cosmology in $d+1$ spacetime dimensions
\be
ds^2=-dt^2+a^2(t)(dx_1^2+\dots dx_d^2)
\label{one}
\ee
In $d=3$ we have $a(t)\sim t^{1\over 2}$ for a radiation dominated universe, $a(t)\sim t^{2\over 3}$ for a dust dominated universe. 

We write $\vec r = (x_1,\dots x_d)$. Consider the point $ \vec r =0$ at time $t_0$. The cosmology has a horizon; i.e.,  there is a radius $|\vec r|=r_p$ from outside which no signal has yet reached the point $\vec r=0$. The question is: should the Hilbert space at time $t_0$ describing the physics of the point $\vec r=0$ contain the degrees of freedom at $|\vec r|>r_p$?

This question may sound like an academic one, since we do not lose anything by letting the Hilbert space include all degrees of freedom at $0\le |\vec r| <\infty$. But the {\it spirit} of quantum mechanics suggests that only what can be measured should be relevant. The degrees of freedom at $|\vec r|>r_p$ cannot be measured by a person at $\vec r=0$ at time $t_0$, so one might hope that there is a way to set things up so that only the degrees of freedom inside the horizon are relevant.

{\it But quantum mechanics is not set up this way !} In quantum theory we have just {\it one} Hilbert space, and this Hilbert space includes all the degrees of freedom in the region $0\le |\vec r| <\infty$. There have been attempts to define theories with many overlapping Hilbert spaces, but it is probably fair to say that at present we do not have a well accepted and standard formulation of quantum theory along such lines. 

The initial state at $t=0$ is not determined by any principle in quantum theory, though one might argue in favor of special states like the Hartle-Hawking state obtained by tunneling from a Euclidean space without any past boundary. Due to the philosophical questions arising from the arbitrariness of the initial state, it is helpful to recast the problem in terms of the `big crunch' rather than the big bang. We assume that our theory is invariant under CPT. Thus we can change $t\rightarrow -t$, and look at a situation where the universe is contracting. On an initial time slice, which we will label $t=-t_0$ we take a homogenous matter distribution, either radiation or dust. The universe contracts, reaching a big crunch singularity at time $t=0$, which is a finite proper time in the future of $t=-t_0$. On the slice $t=-t_0$ there is a value of $|\vec r|=r_p$ such that the point at $\vec r=0$ will not be able to communicate with points at $|\vec r|>r_p$ before the big crunch is encountered. More generally, we can find  patches of space on the slice $t=-t_0$ such that the points in one patch will never be able to communicate with or receive communications from the points in the other patch, before the universe ends at the big crunch. The {\it spirit} on quantum theory suggests that the Hilbert space relevant to one such patch should not contain the degrees of freedom in the other patch. But as we have noted above, the standard formulation of quantum theory is not set up to have different Hilbert spaces for the two patches.

We noted above that this question may sound like an academic one, since there is no {\it contradiction} if the Hilbert space has some degrees of freedom that a given observer cannot interact with. The reason we raise the issue here, however, is that the lesson from black holes will suggest that  
for suitable initial conditions we find a resolution of this vexing situation. For these initial conditions we will argue that the degrees of freedom on the spacelike slice automatically partition themselves into  chunks of size $\sim r_p$, where $r_p$ gets smaller and smaller as we approach the big crunch. 

Black holes are characterized by an outer horizon, from inside which signals cannot escape to infinity. For cosmology, we have multiple notions of the horizon. Before proceeding, we recall the computation of the particle horizon $R_p$ or our example of a flat cosmology (\ref{one}).  A light ray heading towards $\vec r=0$ satisfies
\be
dt=-a(t) dr
\ee
Thus the comoving coordinate elapsed along the path of the light ray since the start of the universe is given by
\be
r_p=\int_{t'=0}^t {dt' \over a(t')}
\ee
For a dust cosmology we have
\be
a(t)=a_0t^{2\over d}
\ee
This gives
\be
r_p={1\over a_0}{d\over d-2} t^{{(d-2)\over d}}
\ee
This corresponds a proper distance of
\be
R_p=a(t) r_p={d\over d-2}t
\ee
which is the `particle horizon' for this cosmology. While the philosophical question noted above is phrased in terms of $R_p$, we will see below that the contradiction we will pose below will be in terms of a slightly different horizon: one defined in terms of the convergence of null rays. 

\section{A sharp puzzle}\label{sectwo}

We now turn to a more concrete problem: one which arises by relating cosmology to black hole physics. Take again the cosmology (\ref{one}), and for concreteness let the matter be dust. We will now relate this cosmological setting to the black hole problem; this relation was noted in \cite{relate}.

Consider a homogenous ball of dust with mass $M$ and radius $R$.  We start with
\be
R=R_0>{2GM}\equiv R_h
\ee
and allow the ball to collapse under its own gravity. In classical general relativity, the  ball will pass through the radius
\be
R=R_h
\ee
and continue collapsing towards a singularity. A horizon is created at $R=R_h$. Hawking pairs are created at this horizon, and this pair creation leads to the information paradox \cite{mat-hawking}. 

As we will see in more detail below, the information paradox can be turned into a sharp statement: if evolution around this horizon is like evolution in vacuum spacetime to leading order (as implied by the semiclassical picture), then we must have remnants or information loss \cite{mat-cern}. We  also note that string allows neither of these two possibilities. The way the information puzzle is resolved is that a horizon never forms: the collapsing ball tunnels into a linear superposition of `fuzzball' states, each of which is a horizon sized quantum ball with no horizon.

\begin{figure}
\begin{center}
 \includegraphics[scale=.8] {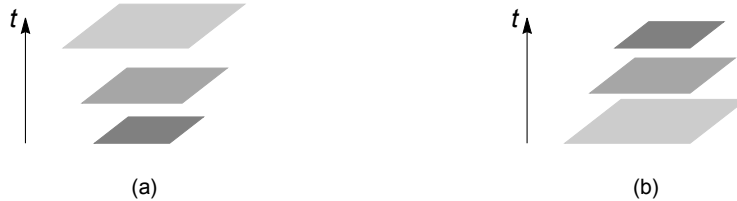}
\end{center}
\caption{(a) An expanding universe can be mapped by CPT to  (b) a collapsing universe. } 
\label{f2}
\end{figure}

Now we relate this description to cosmology. We proceed in steps, making some plausible assumptions at each step:

\bigskip

(i) Consider a slice of the cosmology (\ref{one}) at $t=t_0$. We reverse the direction of time $t\rightarrow -t$ so that the cosmology is heading towards a big crunch (fig.\ref{f2}).

\bigskip

(ii) We mark out  a ball of proper radius $R_b$ around the point $\vec r=0$. We imagine that a thin shell 
\be
R_b<R<R_b+\delta, ~~~\delta \ll R_b
\ee
 is `empty'; i.e. is a vacuum region. This assumption allows us to make a clean separation between the ball $R<R_b$ and the region $R>R_b$. It should not be needed however for our dust cosmology, as there is no direct contact between neighboring dust particles anyway; thus  there is no `pressure' from the region $R>R_b$ onto the surface $R=R_b$.
 
 \bigskip
 
 (iii) We assume that the Birkoff theorem holds; i.e., for the spherically symmetric situation that we have, the region $R>R_b+\delta$ has no influence on the evolution inside the region $R<R_b$.\footnote{Arguments using the Birkoff theorem in cosmology have also been made in \cite{starkman}.} Thus we can replace the dust cosmology in the region $R>R_b+\delta$ with a vacuum region; i.e., asymptotically flat spacetime. Then the dynamics of this collapsing ball becomes identical to the collapsing ball that we had in the black hole problem. 
 
 \bigskip
 
 \begin{figure}[h]
\begin{center}
 \includegraphics[scale=1] {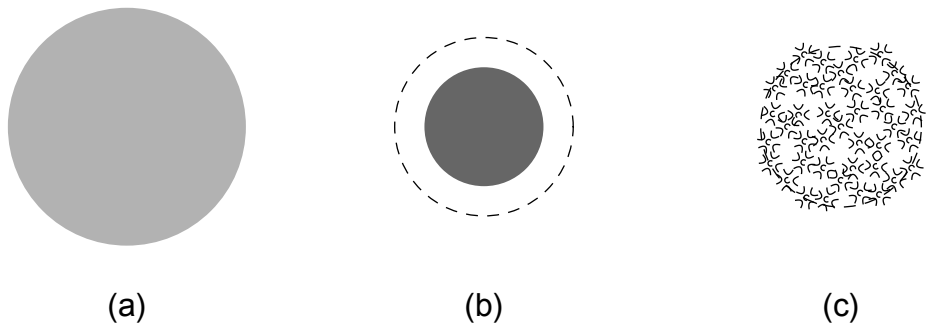}
\end{center}
\caption{ (a) A collapsing dust ball (b) The ball should never reach the semiclassically expected configuration (b), else we cannot solve the information puzzle (c) In string theory fuzzballs form instead when the ball reaches horizon radius. } 
\label{f1}
\end{figure}

 (iv) For the black hole, we had argued that the dust ball never collapses through its horizon; the ball tunnels to fuzzballs when its radius shrinks close to the value $R=R_h$.
 We should therefore expect that in the big crunch, there can never be a region of normal dust  which has
 \be
 R<R_h=2GM
 \label{two}
 \ee
 Here $M$ is defined by connecting the ball $R<R_h$ to asymptotically flat spacetime as in (iii) above. Thus we see a sharp departure from the semiclassical picture, where the dust would simply pass through configurations of higher and higher density (while remaining dust) till it reached infinite density at the big crunch (fig.\ref{f1}).
 
 \bigskip

 (v) We can reverse time $t\rightarrow -t$ to address the big bang. Since string theory is invariant under such a change of time, we argue that in the big bang we can also never have a region satisfying (\ref{two}); i.e., where the dust is inside its own horizon radius.
 
 \bigskip
 
There is nothing in the above argument that says that one has to be close to the big bang or the big crunch; thus the argument should be equally applicable to the universe today. It might seem that the argument is saying that the universe we see should not extend past the cosmological horizon today. If this was all that the argument implied, then there would be no sharp contradiction: we have no direct evidence of matter beyond today's cosmological horizon since we cannot see past this horizon. 
But the argument actually has a stronger implication: at {\it no} time in the past should there have been a region of normal dust which was inside its own horizon. 

This poses a sharp puzzle. Let the time today since the big bang be $t_1$. At times 
$t_2<t_1$, the horizon was smaller. Thus we can find a $t_2$ early enough when the matter was still dust, but where the horizon radius 
$R_{h_2}$ was much smaller than the horizon radius $R_{h_1}$ today:
\be
R_{h_2}\ll R_{h_1}
\ee
Then we expect that there are many patches in the sky today, all within our present horizon $R_{h_1}$, but  which represent regions that had radius $R_{h_2}$ at the earlier time $t_2$. These regions have all smoothly merged to make the interior of our horizon today. It would therefore seem that at time $t_2$ we could have a region of radius
\be
R>R_{h_2}
\ee
without encountering any problem. This would however be in contradiction with our argument leading to the impossibility of (\ref{two}). 

To summarize, our sharp puzzle is as follows. In the traditional picture of cosmology, we have an infinite  homogenous spatial slice. Thus we can always consider a ball of sufficiently large  $R$  so that the mass $M$  inside the ball satisfies $2GM>R$. But if we assume Birkoff's theorem, then we can replace the outside of this ball with empty, asymptotically flat spacetime. This maps the situation to the black hole problem, where we get a dust ball which has collapsed through its horizon; i.e., to a radius $R<R_h=2GM$. But once a black hole horizon forms, there is no resolution to the information puzzle in string theory.

For completeness we review the computation of the radius $R_h$ for the flat cosmology (\ref{one}) in the Appendix.

\section{Two different initial conditions}\label{secc}

In semiclassical gravity we can consider two cases for the dust cosmology:

\b

(C1) As an infinite universe

\b

(C2) As a finite dust ball of radius $R_{max}$; if $R_{max}$ is large and we look at the physics near the center of the ball, then we should not be able to distinguish this case from the case C1 of the infinite universe. 

\b

We will  argue that with the full theory of quantum gravity, the two cases C1 and C2 differ in their full quantum gravity wavefunctional $\Psi$ in an essential way: the virtual fluctuations of fuzzballs are very different in the two cases, and this leads to different evolutions for the full quantum gravity wavefunctional.

\section{Three lessons from black holes}\label{secfour}

To develop our proposal for the nature of the cosmological wavefunctional $\Psi$, we draw on the lessons that we have learnt from three issues with black holes. These three issues are (a) the information paradox  (b) the causality constraint Issue and (c) the conjecture of fuzzball complementarity. Here (a) will tell us that black hole microstates are fuzzballs.  Issue (b) will lead us to a picture of virtual fluctuations of these fuzzballs. The conjecture (c) will give a possible way for an effective semiclassical evolution to arise in case C2 (we will have a direct path to semiclassical evolution in case C1).  

We will then put these lessons together with what we see in the sky to get constraints on how the cosmological wavefunctional should behave in cases C1 and C2.

\subsection{The information paradox and fuzzballs}

Before we make our proposal about the nature of the cosmological wavefunctional $\Psi$, let us recall how we get forced to  the fuzzball paradigm in order to resolve the black hole information paradox. 
We will in fact see that that is a close relation between our issues with cosmology  and the black hole information paradox. 

The information paradox has, of course, been known since 1975, when Hawking found that the formation and evaporation of black holes appeared to violate quantum unitarity. What is the new ingredient that makes the paradox more relevant now than before to the issue of cosmology? As we will now note, the new ingredient is the small corrections theorem which converts Hawking's argument of 1975 into a rigorous result which is stable against all subleading effects that can modify the semiclassical picture \cite{mat-cern}. This theorem removes the hope of bypassing Hawking's argument by some way of introducing subtle corrections to the  radiation from the hole, since the theorem uses quantum information inequalities to prove that the required information cannot be encoded in this way. Thus the only option left is to modify the semiclassical picture completely, so that there is no traditional horizon around which the spacetime would be in the vacuum state to leading order. In particular we must violate the no-hair theorem to get this altered picture of the black hole. In string theory we indeed find a full set of hair; i.e., one horizon sized `fuzzball' with no horizon for each of the $Exp[S_{bek}]$ states implied by the Bekenstein entropy. With this change in the picture of the black hole, we then argue that there must be a change in our picture of cosmology as well. 

Let us start with Hawking's argument \cite{mat-hawking}. Around the horizon of the black hole we have the creation of particle-antiparticle pairs. For example, we could have an electron that falls into the hole, and  a positron that escapes to infinity. The positron is then a quantum of `Hawking radiation', and this radiation process causes the hole to slowly lose its mass.

We have an equal probability for a positron to fall into the hole and an electron to escape to infinity. In fact the overall state of the created pair is an entangled one between the infalling particle and the outgoing one. Calling the electron $0$ and the positron $1$, the state of the created pair has the schematic form
\be
\psi_{pair}={1\over \sqrt{2}}\left( 0_{in}1_{out}+1_{in}0_{out}\right ) \equiv {1\over \sqrt{2}} (00+11)
\label{wten}
\ee
The entanglement entropy of the outgoing particle with the infalling one is then
\be
S_{ent}=\ln 2
\label{wel}
\ee
In Hawking's leading order picture, each created pair is independent of all other created pairs. After $N$ steps of emission, then entanglement of the radiation outside with the infalling quanta is 
 \be
 S_{ent}=N\ln 2
 \ee
 If the black hole evaporates away, then we are left with radiation that is entangled, but there is nothing that it is entangled {\it with}. Such radiation cannot be attributed any quantum state, and can only be described by a density matrix which gives the probabilities (but not the phases) of different possible state. We started with a star which had a definite wavefunction, and after forming and evaporating the hole, we are left with radiation that is defined only statistically, by its density matrix. Thus we have violated the unitary evolution of quantum theory \cite{mat-hawking}
 \be
 \Psi_{final}=e^{-i \hat H t} \Psi _{initial}
 \ee
To escape this problem we may conjecture  that the hole not evaporate away but stop its evaporation when it becomes a planck sized remnant. But such planck size remnants must have an infinite degeneracy of states, to be able to describe all the possible states of infalling particles that can result from the evaporation of holes with arbitrarily large initial mass $M$. Such remnants are not allowed in string theory, where we have only a finite number of states below a given mass $m_p$ confined to a given radius $l_p$.

One may therefore hope that Hawking's argument could be invalidated by subleading corrections to the leading semiclassical computation. Hawking had ignored correlations between different emitted quanta, but of course there can be small quantum gravity effects of order $\epsilon\ll 1$ that would create such correlations. The number of emitted quanta $N$ is very large
\be
N\sim \left ( {M\over m_p}\right )^2  \sim S_{bek}
\ee
where $S_{bek}={A/4G}$ is the Bekenstein entropy of the hole. Thus there might be an exponentially large number of small correction terms to the overall state $\Psi$ describing  the radiation and the remaining hole. Even though the correction to each emitted pair is small, it might be that these small corrections cumulate (by the end of the evaporation process) to remove the net entanglement between the radiation and the hole. In that case the hole can vanish away, leaving a pure, unentangled state of the radiation.

But the small correction theorem removes this possibility. Let $S_{ent,n}$ be the entanglement of the emitted radiation with the remaining hole after $n$ steps of emission. Using the strong subadditivity of quantum entanglement entropy, it was shown that \cite{mat-cern}
\be
S_{ent, n+1}> S_{ent,n}+\ln 2 -2\epsilon
\label{weightq}
\ee
where $\epsilon$ bounds the correction at each step of the emission, We see that the number of emitted quanta $N$ does not appear in this inequality,, so we cannot use the largeness of $N$ to offset the smallness of $\epsilon$ and resolve the Hawking puzzle. 

We can summarize this conclusion as follows: to resolve the information paradox, we need {\it order unity} corrections to the low energy dynamics at the horizon. Fuzzballs provide just such a resolution, and so we will accept the fuzzball paradigm as the resolution of the information paradox in what follows. 

\subsection{Causality}

Given that the black hole has to be replaced by a fuzzball, we can ask: {\it when} exactly does such a fuzzball have to form? Consider a ball of dust that is collapsing to make a hole. Should the ball pass smoothly through its horizon, as suggested by semiclassical physics, and then evolve into a fuzzball later? Or should the collapsed be halted outside the horizon location through a process of fuzzball formation? We will argue that the latter must be the case; for details, see \cite{mat-causality}.

Consider the classical black hole metric with horizon radius $R_h$. No timelike or null curve can escape from the region $R<R_h$ to the region $R>R_h$; so in classical general relativity nothing can come out of the horizon. Thus in classical physics we can make a black hole but not `unmake it'. 
One might think that quantum fluctuations can bring information from inside the horizon to the outside. But if we study `quantum fields in curved space', then we still find that information does not travel faster than light. The  perturbative fluctuations of gravity can be studied using a quantum field $h_{\mu\nu}$ on the background (\ref{one}), so metric fluctuations also travel inside the light cone. Nonperturbative quantum gravity is less well understood, but a simple example is provided by bubble nucleation in cosmology. The bubble wall  travels below the speed of light, so even with such nonperturbative effects  we cannot send signals outside the light cone by triggering the creation of such bubbles. String theory has extended objects like strings, but these also do not allow faster than light signals: if we perturb one end of a string, the disturbance travels along the string at a speed less than or equal to the speed of light. In short, we do not know of any effect in physics that can send signals outside the light cone.

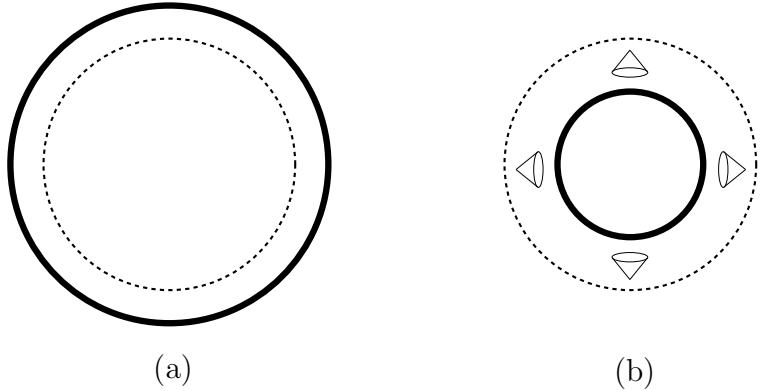
\begin{figure}
\hskip 1 in
\begin{tikzpicture}[y=0.80pt, x=.80pt, yscale=-.156000000, xscale=.156000000, inner sep=0pt, outer sep=0pt]
  \begin{scope}[shift={(0,-48.57141)}]
    \path[draw=black,dash pattern=on 1.60pt off 1.60pt,line join=miter,line
      cap=butt,miter limit=4.00,even odd rule,line width=0.800pt]
      (368.5714,580.9336) circle (10.7244cm);
    \path[draw=black,line join=miter,line cap=butt,miter limit=4.00,even odd
      rule,line width=2.400pt] (368.5714,580.9336) circle (6.2089cm);
    \path[draw=black,line join=miter,line cap=butt,miter limit=4.00,even odd
      rule,line width=0.283pt] (367.1428,302.1688) ellipse (1.5232cm and 0.3993cm);
    \path[draw=black,line join=miter,line cap=butt,miter limit=4.00,even odd
      rule,line width=0.283pt] (313.9304,300.1467) -- (367.5256,234.0937) --
      (367.5256,234.0937) -- (367.5256,234.0937) -- (367.5256,234.0937);
    \path[draw=black,line join=miter,line cap=butt,miter limit=4.00,even odd
      rule,line width=0.283pt] (421.0006,300.8207) -- (367.4054,234.7677) --
      (367.4054,234.7677) -- (367.4054,234.7677) -- (367.4054,234.7677);
    \path[xscale=1.000,yscale=-1.000,draw=black,line join=miter,line cap=butt,miter
      limit=4.00,even odd rule,line width=0.283pt] (367.1428,-862.5269) ellipse
      (1.5232cm and 0.3993cm);
    \path[draw=black,line join=miter,line cap=butt,miter limit=4.00,even odd
      rule,line width=0.283pt] (313.9304,864.5489) -- (367.5256,930.6019) --
      (367.5256,930.6019) -- (367.5256,930.6019) -- (367.5256,930.6019);
    \path[draw=black,line join=miter,line cap=butt,miter limit=4.00,even odd
      rule,line width=0.283pt] (421.0006,863.8749) -- (367.4054,929.9279) --
      (367.4054,929.9279) -- (367.4054,929.9279) -- (367.4054,929.9279);
    \path[cm={{0.0,-1.0,1.0,0.0,(0.0,0.0)}},draw=black,line join=miter,line
      cap=butt,miter limit=4.00,even odd rule,line width=0.283pt]
      (-596.6335,89.8208) ellipse (1.5232cm and 0.3993cm);
    \path[draw=black,line join=miter,line cap=butt,miter limit=4.00,even odd
      rule,line width=0.283pt] (87.7988,649.8459) -- (21.7458,596.2507) --
      (21.7458,596.2507) -- (21.7458,596.2507) -- (21.7458,596.2507);
    \path[draw=black,line join=miter,line cap=butt,miter limit=4.00,even odd
      rule,line width=0.283pt] (88.4728,542.7757) -- (22.4198,596.3709) --
      (22.4198,596.3709) -- (22.4198,596.3709) -- (22.4198,596.3709);
    \path[cm={{0.0,-1.0,-1.0,0.0,(0.0,0.0)}},draw=black,line join=miter,line
      cap=butt,miter limit=4.00,even odd rule,line width=0.283pt]
      (-596.6335,-650.1790) ellipse (1.5232cm and 0.3993cm);
    \path[draw=black,line join=miter,line cap=butt,miter limit=4.00,even odd
      rule,line width=0.283pt] (652.2010,649.8459) -- (718.2540,596.2507) --
      (718.2540,596.2507) -- (718.2540,596.2507) -- (718.2540,596.2507);
    \path[draw=black,line join=miter,line cap=butt,miter limit=4.00,even odd
      rule,line width=0.283pt] (651.5270,542.7757) -- (717.5800,596.3709) --
      (717.5800,596.3709) -- (717.5800,596.3709) -- (717.5800,596.3709);
  \end{scope}
  \begin{scope}[shift={(2.85714,-40.0)}]
    \path[draw=black,dash pattern=on 1.60pt off 1.60pt,line join=miter,line
      cap=butt,miter limit=4.00,even odd rule,line width=0.800pt]
      (-1031.4286,572.3622) circle (10.7244cm);
    \path[draw=black,line join=miter,line cap=butt,miter limit=4.00,even odd
      rule,line width=2.400pt] (-1031.4286,572.3622) circle (13.5467cm);
  \end{scope}
  \path[fill=black,line join=miter,line cap=butt,line width=0.800pt]
    (-1077.4313,1204.3622) node[above right] (text4470) {(a)};
  \path[fill=black,line join=miter,line cap=butt,line width=0.800pt]
    (319.0280,1212.9629) node[above right] (text4474) {(b)};

\end{tikzpicture}

\caption{(a) A shell of mass $M$ is collapsing towards its horizon. (b) If the shell passes through its horizon, then the information it carries is trapped inside the horizon due to the structure of light cones.} \label{fs}

\end{figure}

Thus if we first allow the ball of dust to fall through its horizon (fig.\ref{fs}), then we cannot allow its information to escape by any means that are known in string theory,. We must therefore halt the collapse before the horizon forms. In \cite{mat-causality} a picture was given where tunneling into fuzzballs leads to exactly such a halt to the collapse process: the infalling ball tunnels into a linear combination of fuzzball just outside the location where it would have created a horizon. 

To summarize, what we have learnt from string theory is that {\it matter never gets compressed to a point where it can fit inside its own horizon}. As we try to compress matter to the point where it might form a horizon, we find that it tunnels into fuzzballs, and the horizon never forms. 

\subsection{Fuzzball complementarity}

The fuzzball is a very quantum gravitational object, and the interior of the fuzzball is not
anywhere close to a vacuum spacetime. Can we then recover, in any approximation, the traditional classical picture of infall into the black hole interior?

It was argued in \cite{fuzzcomp,beyond,mat-causality} that such a picture of infall could be recovered in an {\it approximate} way in a {\it dual } description.\footnote{This duality is similar to AdS/CFT duality \cite{adscft} but also different in an essential way. While AdS/CFT duality is an exact map, fuzzball complementarity is an approximate emergent duality valid only for infalling observers with energies $E\gg T$.  For details see \cite{mat-causality}.} The infalling dust ball transitions into a linear combinations of fuzzballs, and these fuzzballs do not have a smooth vacuum region around the horizon radius. But the ball does not just transition to fuzzballs: the fuzzball state continues to evolve in the {\it space of all fuzzball solutions}. This is a very large space of dimension $Exp[S_{bek}]$; we will term it $H_F$ The conjecture of fuzzball complementarity says that {\it short time and short distance evolution in this large space $H_F$ can be approximately mapped to evolution in the traditional semiclassical black hole}. Here the phrase `short time' means that this approximation isn valid only for times of order the infall time in the classical hole. The phrase `short distance' means that the region over which we can get the dual description should have a size $L\ll R_h$.

 An explicit model for this evolution exhibiting fuzzball complementarity was given in \cite{mat-bitmodel}.

\section{The proposal}\label{secp}

We will propose a model for the behavior of fuzzballs, by assuming the following properties of fuzzballs:

\b

(F1): The microstates of black holes are fuzzballs, which are horizon sized quantum objects with no horizon or singularity. Simple fuzzballs can be given a description in semiclassical gravity, and from such constructions we deduce the existence of generic  microstates which do not have a goos semiclassical  description. There are 
\be
{\cal N}\sim e^{S_{bek}(M)}
\label{bbthree}
\ee
fuzzball states with mass $M$. 

\b

(F2): The fuzzballs are rather  incompressible objects (for a toy model of a fuzzball, see \cite{model}). More precisely, they are compression resistant (no structure can be completely incompressible in a theory where signals cannot propagate faster than the speed of light).  If we try to squeeze a fuzzball, then its energy rises rapidly. This extra energy will typically cause the fuzzball state $F_i(M)$ at energy $M$ to change so that it becomes a linear combination of other fuzzball states $F_j(M+\delta M)$ at a higher energy $M+\delta M$:
\be
F_i(M)\r \sum_j C_{ij} F_j(M+\delta M)
\ee
Thus we say that fuzzballs are `extended compression resistant objects'. Note that a string loop can be an extended object, but it is not incompressible in the same way: even if the string is carrying vibrations, it can be compressed with a much lower cost in fractional energy. With fuzzballs, we cannot fit the number ${\cal N}$ of fuzzball states (\ref{bbthree}) in a region of size smaller than $\approx 2M+l_p$, so the generic state must necessarily be incompressible in the sense mentioned above (fig.\ref{fig2qq}(d)).

\b

\begin{figure}
\hskip 1 true in \includegraphics[scale=.12] {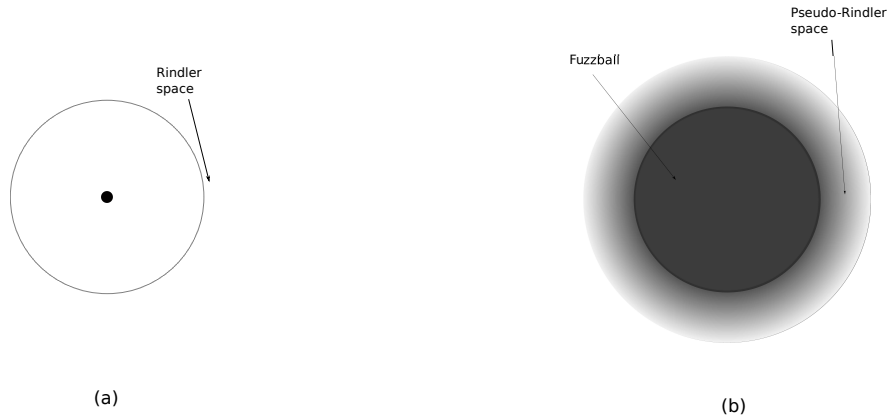}
\caption{\label{figrindlercausalityp} (a) In the traditional hole, the region just {\it outside} the horizon is Rindler space, which is just a part of Minkowski space. (b) The dark circle is the fuzzball. The region outside the fuzzball has extra  vacuum fluctuations that correspond to the fuzzball of mass $M$ fluctuating to a fuzzball of mass $M_f>M$. These virtual fuzzballs are depicted by the shaded region outside the fuzzball. Because of altered vacuum fluctuations, the region near the fuzzball boundary is  termed pseudo-Rindler space.}
\end{figure}

(F3): Suppose we have a black hole state of mass $M$;  i.e., a fuzzball $F_i(M)$. Then there are virtual fluctuations about this state describing fuzzball states $F_j(M')$ with higher energy $M'>M$
\be
F_i(M) \r F_j(M')
\label{bbone}
\ee
The probability for  fluctuating to any one fuzzball state $F_j(M')$ is small for $M'-M\gg m_p$, but the {\it number} of fuzzball states with  mass $M'$ is very {\it large}\cite{tunnel,kraus,puhm}
\be
{\cal N}'\sim e^{S_{bek}(M')}
\ee
so the overall effect of these virtual fluctuations is significant, rather than a small quantum correction to the dynamics (fig.\ref{figrindlercausalityp}). 

\b

(F4): The effect of these virtual fluctuations is to give spacetime another parameter -- a `depth'. The fuzzball $F_i(M)$ itself typically ends at a radius $r_f\approx 2M+l_p$, i..e, just outside the location where the horizon would be in the traditional hole. The multipole moments of the fuzzball structure die off rapidly at $r>r_f$, so one might think that spacetime at $r>r_f$ is essential the Schwarzschild geometry. But in \cite{mat-causality} it was argued that the {\it virtual fluctuations} (\ref{bbone}) are present in this region $r>r_f$ and these fluctuations make this region different from a patch of spacetime near infinity. 
The alteration can be schematically described by modeling spacetime as a sea, whose depth is infinity at $r\r\infty$ and reduces gradually to zero at $r\approx 2M+l_p$. 

Such a picture of a varying depth sea was found in the description of the $0+1$ dimensional matrix model, where the eigenvalues filled a fermi sea whose dynamics gave rise to 1+1 dimensional spacetime \cite{dasjevicki}. Suppose the depth pf the sea at a given point is $d$. A wave with amplitude $A\ll d$ will not notice that $d$ is finite; it will move as a ripple on the surface of the sea almost as if the sea had infinite depth. The motion of such a ripple gives the motion of the massless tachyon field in a 1+1 dimensional dilaton gravity background. On the other hand if $A\gtrsim d$, then the wave will feel the bottom of the sea, and distort correspondingly; in fact it can fold over itself, at which point it will describe not a coherent state of the tachyon but instead a complicated quantum state of the tachyon with large dispersions \cite{folds}. 

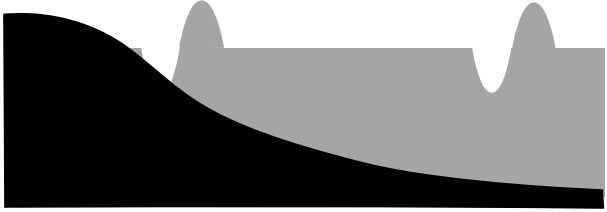
\begin{figure}

\definecolor{ca5a5a5}{RGB}{165,165,165}
\definecolor{cffffff}{RGB}{255,255,255}

\begin{tikzpicture}[y=0.80pt, x=0.80pt, yscale=-.17000000, xscale=.17000000, inner sep=0pt, outer sep=0pt]
  \path[fill=ca5a5a5,line join=miter,line cap=butt,miter limit=4.00,even odd
    rule,line width=0.000pt] (-708.5714,260.9336) .. controls (-708.5714,260.9336)
    and (-687.9726,126.6479) .. (-646.7750,126.6479) .. controls
    (-605.5773,126.6479) and (-585.7142,258.5357) .. (-585.7142,258.5357) --
    cycle;
  \path[fill=ca5a5a5,miter limit=4.00,draw opacity=0.986,line
    width=0.000pt,rounded corners=0.0000cm] (-1831.4286,252.3622) rectangle
    (-451.4286,666.6479);
  \path[fill=ca5a5a5,line join=miter,line cap=butt,miter limit=4.00,even odd
    rule,line width=0.000pt] (-1634.2857,255.2193) .. controls
    (-1634.2857,255.2193) and (-1613.2078,120.9336) .. (-1571.0522,120.9336) ..
    controls (-1528.8964,120.9336) and (-1508.5713,252.8214) ..
    (-1508.5713,252.8214) -- cycle;
  \path[fill=cffffff,line join=miter,line cap=butt,miter limit=4.00,even odd
    rule,line width=0.000pt] (-820.0000,243.7908) .. controls (-820.0000,243.7908)
    and (-801.3173,378.0765) .. (-763.9521,378.0765) .. controls
    (-726.5867,378.0765) and (-708.5713,246.1887) .. (-708.5713,246.1887) --
    cycle;
  \path[fill=cffffff,line join=miter,line cap=butt,miter limit=4.00,even odd
    rule,line width=0.000pt] (-1739.9999,246.6479) .. controls
    (-1739.9999,246.6479) and (-1721.7963,380.9336) .. (-1685.3891,380.9336) ..
    controls (-1648.9819,380.9336) and (-1631.4284,249.0458) ..
    (-1631.4284,249.0458) -- cycle;
  \path[draw=black,fill=black,line join=miter,line cap=butt,even odd rule,line
    width=0.800pt] (-451.4286,698.0765) .. controls (-1000.0000,692.3622) and
    (-1567.1282,693.3293) .. (-2117.1428,695.2193) -- (-2120.0000,160.9336) ..
    controls (-2120.0000,160.9336) and (-1934.2857,135.2193) ..
    (-1774.2857,255.2193) .. controls (-1614.2857,375.2193) and
    (-1597.1428,455.2193) .. (-1117.1428,575.2193) .. controls
    (-881.0970,634.2308) and (-454.2857,649.5050) .. (-454.2857,649.5050);

\end{tikzpicture}

\caption{  The black region is land, while the grey region is water.  A wave of  travels freely when its amplitude is much less than the depth of the water, but will suffer nontrivial deformation when the amplitude becomes comparable to the depth of the water.}\label{fig3qq}

\end{figure}

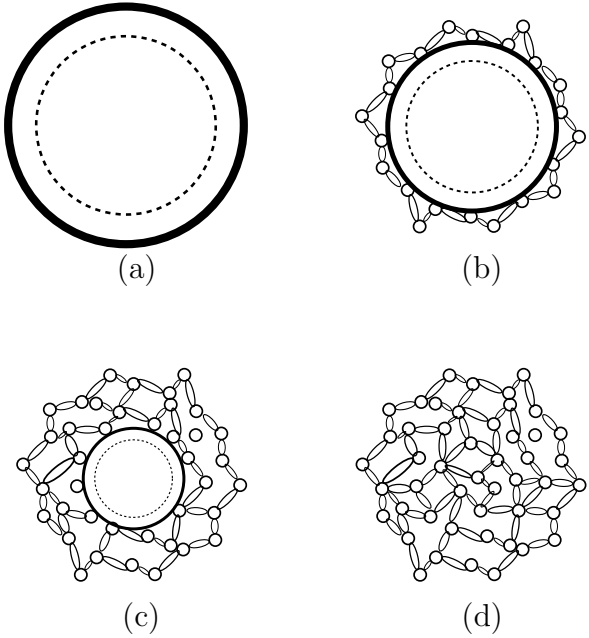
\begin{figure}

\begin{tikzpicture}[y=0.80pt, x=0.80pt, yscale=-.09000000, xscale=.09000000, inner sep=0pt, outer sep=0pt]
  \begin{scope}[shift={(-471.42857,28.57143)}]
    \path[draw=black,dash pattern=on 1.97pt off 1.97pt,line join=miter,line
      cap=butt,miter limit=4.00,even odd rule,line width=0.986pt]
      (-971.4286,-190.4949) ellipse (13.2611cm and 13.1805cm);
    \path[draw=black,line join=miter,line cap=butt,miter limit=4.00,even odd
      rule,line width=3.098pt] (-971.4286,-190.4949) ellipse (17.4048cm and
      17.5661cm);
  \end{scope}
  \path[fill=black,line join=miter,line cap=butt,line width=0.800pt]
    (-1491.7170,682.8970) node[above right] (text4470) {(a)};
  \path[fill=black,line join=miter,line cap=butt,line width=0.800pt]
    (317.5994,682.9312) node[above right] (text4474) {(b)};
    \path[draw=black,dash pattern=on 1.45pt off 1.45pt,line join=miter,line
      cap=butt,miter limit=4.00,even odd rule,line width=0.725pt]
      (375.7143,-154.7806) circle (9.7178cm);
    \path[draw=black,line join=miter,line cap=butt,miter limit=4.00,even odd
      rule,line width=1.888pt] (375.7143,-154.7806) circle (12.4615cm);
    \path[draw=black,line join=miter,line cap=butt,miter limit=4.00,even odd
      rule,line width=0.640pt] (254.2856,-679.0664) circle (0.9031cm);
    \path[cm={{0.70711,-0.70711,0.70711,0.70711,(0.0,0.0)}},draw=black,line
      join=miter,line cap=butt,miter limit=4.00,even odd rule,line width=0.480pt]
      (556.9441,-302.3856) ellipse (2.0320cm and 0.5080cm);
    \path[cm={{0.97557,-0.21967,0.14792,0.989,(0.0,0.0)}},draw=black,line
      join=miter,line cap=butt,miter limit=4.00,even odd rule,line width=0.414pt]
      (98.2324,-503.0207) ellipse (1.5210cm and 0.5039cm);
    \path[cm={{-0.05183,-0.99866,0.98513,-0.17183,(0.0,0.0)}},draw=black,line
      join=miter,line cap=butt,miter limit=4.00,even odd rule,line width=0.356pt]
      (433.3398,-38.1080) ellipse (1.1146cm and 0.5086cm);
    \path[cm={{-0.61729,-0.78673,0.86996,-0.49312,(0.0,0.0)}},draw=black,line
      join=miter,line cap=butt,miter limit=4.00,even odd rule,line width=0.401pt]
      (206.1191,-29.4514) ellipse (1.6452cm and 0.4376cm);
    \path[draw=black,line join=miter,line cap=butt,miter limit=4.00,even odd
      rule,line width=0.640pt] (105.7143,-539.0664) circle (0.9031cm);
    \path[draw=black,line join=miter,line cap=butt,miter limit=4.00,even odd
      rule,line width=0.640pt] (-62.8572,-499.0664) circle (0.9031cm);
    \path[draw=black,line join=miter,line cap=butt,miter limit=4.00,even odd
      rule,line width=0.640pt] (-202.8572,-210.4950) circle (0.9031cm);
    \path[draw=black,line join=miter,line cap=butt,miter limit=4.00,even odd
      rule,line width=0.640pt] (-57.1429,-356.2093) circle (0.9031cm);
    \path[cm={{0.70711,-0.70711,0.70711,0.70711,(0.0,0.0)}},draw=black,line
      join=miter,line cap=butt,miter limit=4.00,even odd rule,line width=0.480pt]
      (112.4770,-290.2637) ellipse (2.0320cm and 0.5080cm);
    \path[cm={{-0.66187,-0.74962,0.79591,-0.60542,(0.0,0.0)}},draw=black,line
      join=miter,line cap=butt,miter limit=4.00,even odd rule,line width=0.414pt]
      (-70.8148,-107.3516) ellipse (1.5210cm and 0.5039cm);
    \path[cm={{0.05183,0.99866,-0.98513,0.17183,(0.0,0.0)}},draw=black,line
      join=miter,line cap=butt,miter limit=4.00,even odd rule,line width=0.356pt]
      (-28.6468,97.1022) ellipse (1.1146cm and 0.5086cm);
    \path[scale=-1.000,draw=black,line join=miter,line cap=butt,miter
      limit=4.00,even odd rule,line width=0.640pt] (-37.1428,-178.0765) circle
      (0.9031cm);
    \path[scale=-1.000,draw=black,line join=miter,line cap=butt,miter
      limit=4.00,even odd rule,line width=0.640pt] (94.2856,-60.9338) circle
      (0.9031cm);
    \path[scale=-1.000,draw=black,line join=miter,line cap=butt,miter
      limit=4.00,even odd rule,line width=0.640pt] (100.0000,81.9235) circle
      (0.9031cm);
    \path[draw=black,line join=miter,line cap=butt,miter limit=4.00,even odd
      rule,line width=0.640pt] (99.9999,369.5050) circle (0.9031cm);
    \path[cm={{-0.23861,-0.97112,0.97112,-0.23861,(0.0,0.0)}},draw=black,line
      join=miter,line cap=butt,miter limit=4.00,even odd rule,line width=0.480pt]
      (-280.8569,1.6033) ellipse (2.0320cm and 0.5080cm);
    \path[cm={{0.63996,-0.7684,0.4865,0.87368,(0.0,0.0)}},draw=black,line
      join=miter,line cap=butt,miter limit=4.00,even odd rule,line width=0.228pt]
      (-35.5051,334.4735) ellipse (0.7970cm and 0.2911cm);
    \path[cm={{-0.7825,-0.62265,0.87864,-0.47748,(0.0,0.0)}},draw=black,line
      join=miter,line cap=butt,miter limit=4.00,even odd rule,line width=0.257pt]
      (470.1735,773.1666) ellipse (0.9230cm and 0.3201cm);
    \path[draw=black,line join=miter,line cap=butt,miter limit=4.00,even odd
      rule,line width=0.640pt] (182.8571,278.0766) circle (0.9031cm);
    \path[draw=black,line join=miter,line cap=butt,miter limit=4.00,even odd
      rule,line width=0.640pt] (377.1429,-633.3521) circle (0.9031cm);
    \path[draw=black,line join=miter,line cap=butt,miter limit=4.00,even odd
      rule,line width=0.640pt] (368.5715,318.0765) circle (0.9031cm);
    \path[cm={{0.89985,0.43621,-0.43621,0.89985,(0.0,0.0)}},draw=black,line
      join=miter,line cap=butt,miter limit=4.00,even odd rule,line width=0.480pt]
      (386.9658,156.3681) ellipse (2.0320cm and 0.5080cm);
    \path[scale=-1.000,draw=black,line join=miter,line cap=butt,miter
      limit=4.00,even odd rule,line width=0.640pt] (-937.1428,101.9235) circle
      (0.9031cm);
    \path[scale=-1.000,draw=black,line join=miter,line cap=butt,miter
      limit=4.00,even odd rule,line width=0.640pt] (-489.4067,-366.6479) circle
      (0.9031cm);
    \path[cm={{-0.70711,0.70711,-0.70711,-0.70711,(0.0,0.0)}},draw=black,line
      join=miter,line cap=butt,miter limit=4.00,even odd rule,line width=0.480pt]
      (-189.8390,-607.3423) ellipse (2.0320cm and 0.5080cm);
    \path[cm={{-0.97557,0.21967,-0.14792,-0.989,(0.0,0.0)}},draw=black,line
      join=miter,line cap=butt,miter limit=4.00,even odd rule,line width=0.414pt]
      (-685.5795,-361.2249) ellipse (1.5210cm and 0.5039cm);
    \path[cm={{0.05183,0.99866,-0.98513,0.17183,(0.0,0.0)}},draw=black,line
      join=miter,line cap=butt,miter limit=4.00,even odd rule,line width=0.356pt]
      (252.0342,-802.5671) ellipse (1.1146cm and 0.5086cm);
    \path[cm={{0.61676,0.78715,-0.86966,0.49365,(0.0,0.0)}},draw=black,line
      join=miter,line cap=butt,miter limit=4.00,even odd rule,line width=0.327pt]
      (291.4953,-815.9766) ellipse (1.3428cm and 0.3570cm);
    \path[scale=-1.000,draw=black,line join=miter,line cap=butt,miter
      limit=4.00,even odd rule,line width=0.640pt] (-637.9783,-236.6478) circle
      (0.9031cm);
    \path[scale=-1.000,draw=black,line join=miter,line cap=butt,miter
      limit=4.00,even odd rule,line width=0.640pt] (-806.5496,-186.6479) circle
      (0.9031cm);
    \path[scale=-1.000,draw=black,line join=miter,line cap=butt,miter
      limit=4.00,even odd rule,line width=0.640pt] (-800.8352,-43.7908) circle
      (0.9031cm);
    \path[cm={{-0.70711,0.70711,-0.70711,-0.70711,(0.0,0.0)}},draw=black,line
      join=miter,line cap=butt,miter limit=4.00,even odd rule,line width=0.480pt]
      (-634.3062,-595.2205) ellipse (2.0320cm and 0.5080cm);
    \path[cm={{0.66187,0.74962,-0.79591,0.60542,(0.0,0.0)}},draw=black,line
      join=miter,line cap=butt,miter limit=4.00,even odd rule,line width=0.414pt]
      (131.3153,-873.6587) ellipse (1.5210cm and 0.5039cm);
    \path[cm={{-0.05183,-0.99866,0.98513,-0.17183,(0.0,0.0)}},draw=black,line
      join=miter,line cap=butt,miter limit=4.00,even odd rule,line width=0.356pt]
      (152.6587,861.5613) ellipse (1.1146cm and 0.5086cm);
    \path[draw=black,line join=miter,line cap=butt,miter limit=4.00,even odd
      rule,line width=0.640pt] (706.5496,-490.4950) circle (0.9031cm);
    \path[draw=black,line join=miter,line cap=butt,miter limit=4.00,even odd
      rule,line width=0.640pt] (837.9780,-373.3523) circle (0.9031cm);
    \path[draw=black,line join=miter,line cap=butt,miter limit=4.00,even odd
      rule,line width=0.640pt] (843.6924,-230.4950) circle (0.9031cm);
    \path[scale=-1.000,draw=black,line join=miter,line cap=butt,miter
      limit=4.00,even odd rule,line width=0.640pt] (-643.6924,681.9236) circle
      (0.9031cm);
    \path[cm={{0.23861,0.97112,-0.97112,0.23861,(0.0,0.0)}},draw=black,line
      join=miter,line cap=butt,miter limit=4.00,even odd rule,line width=0.480pt]
      (-406.8013,-795.1537) ellipse (2.0320cm and 0.5080cm);
    \path[cm={{-0.63996,0.7684,-0.4865,-0.87368,(0.0,0.0)}},draw=black,line
      join=miter,line cap=butt,miter limit=4.00,even odd rule,line width=0.228pt]
      (-894.8624,-63.7450) ellipse (0.7970cm and 0.2911cm);
    \path[cm={{0.7825,0.62265,-0.87864,0.47748,(0.0,0.0)}},draw=black,line
      join=miter,line cap=butt,miter limit=4.00,even odd rule,line width=0.257pt]
      (557.7043,4.7106) ellipse (0.9230cm and 0.3201cm);
    \path[scale=-1.000,draw=black,line join=miter,line cap=butt,miter
      limit=4.00,even odd rule,line width=0.640pt] (-566.5496,585.6380) circle
      (0.9031cm);
    \path[cm={{-0.89985,-0.43621,0.43621,-0.89985,(0.0,0.0)}},draw=black,line
      join=miter,line cap=butt,miter limit=4.00,even odd rule,line width=0.480pt]
      (-145.9623,761.9019) ellipse (2.0320cm and 0.5080cm);
  \begin{scope}[shift={(0,8.57143)}]
    \path[fill=black,line join=miter,line cap=butt,line width=0.800pt]
      (-1462.4006,2505.7883) node[above right] (text4474-6) {(c)};
      \path[draw=black,dash pattern=on 0.86pt off 0.86pt,line join=miter,line
        cap=butt,miter limit=4.00,even odd rule,line width=0.428pt]
        (-1402.8572,1686.6479) circle (5.7316cm);
      \path[draw=black,line join=miter,line cap=butt,miter limit=4.00,even odd
        rule,line width=1.126pt] (-1402.8572,1686.6479) circle (7.4352cm);
      \path[draw=black,line join=miter,line cap=butt,miter limit=4.00,even odd
        rule,line width=0.640pt] (-1525.7144,1143.7906) circle (0.9031cm);
      \path[cm={{0.70711,-0.70711,0.70711,0.70711,(0.0,0.0)}},draw=black,line
        join=miter,line cap=butt,miter limit=4.00,even odd rule,line width=0.480pt]
        (-1990.6606,-272.0811) ellipse (2.0320cm and 0.5080cm);
      \path[cm={{0.97557,-0.21967,0.14792,0.989,(0.0,0.0)}},draw=black,line
        join=miter,line cap=butt,miter limit=4.00,even odd rule,line width=0.414pt]
        (-1937.2399,887.9988) ellipse (1.5210cm and 0.5039cm);
      \path[cm={{-0.05183,-0.99866,0.98513,-0.17183,(0.0,0.0)}},draw=black,line
        join=miter,line cap=butt,miter limit=4.00,even odd rule,line width=0.356pt]
        (-1067.4927,-1923.9407) ellipse (1.1146cm and 0.5086cm);
      \path[cm={{-0.61729,-0.78673,0.86996,-0.49312,(0.0,0.0)}},draw=black,line
        join=miter,line cap=butt,miter limit=4.00,even odd rule,line width=0.401pt]
        (-509.9290,-2583.6077) ellipse (1.6452cm and 0.4376cm);
      \path[draw=black,line join=miter,line cap=butt,miter limit=4.00,even odd
        rule,line width=0.640pt] (-1674.2858,1283.7908) circle (0.9031cm);
      \path[draw=black,line join=miter,line cap=butt,miter limit=4.00,even odd
        rule,line width=0.640pt] (-1842.8572,1323.7906) circle (0.9031cm);
      \path[draw=black,line join=miter,line cap=butt,miter limit=4.00,even odd
        rule,line width=0.640pt] (-1982.8572,1612.3621) circle (0.9031cm);
      \path[draw=black,line join=miter,line cap=butt,miter limit=4.00,even odd
        rule,line width=0.640pt] (-1837.1429,1466.6478) circle (0.9031cm);
      \path[cm={{0.70711,-0.70711,0.70711,0.70711,(0.0,0.0)}},draw=black,line
        join=miter,line cap=butt,miter limit=4.00,even odd rule,line width=0.480pt]
        (-2435.1277,-259.9591) ellipse (2.0320cm and 0.5080cm);
      \path[cm={{-0.66187,-0.74962,0.79591,-0.60542,(0.0,0.0)}},draw=black,line
        join=miter,line cap=butt,miter limit=4.00,even odd rule,line width=0.414pt]
        (-444.9843,-2654.9565) ellipse (1.5210cm and 0.5039cm);
      \path[cm={{0.05183,0.99866,-0.98513,0.17183,(0.0,0.0)}},draw=black,line
        join=miter,line cap=butt,miter limit=4.00,even odd rule,line width=0.356pt]
        (1472.1857,1982.9349) ellipse (1.1146cm and 0.5086cm);
      \path[scale=-1.000,draw=black,line join=miter,line cap=butt,miter
        limit=4.00,even odd rule,line width=0.640pt] (1742.8572,-2000.9336) circle
        (0.9031cm);
      \path[scale=-1.000,draw=black,line join=miter,line cap=butt,miter
        limit=4.00,even odd rule,line width=0.640pt] (1874.2856,-1883.7909) circle
        (0.9031cm);
      \path[scale=-1.000,draw=black,line join=miter,line cap=butt,miter
        limit=4.00,even odd rule,line width=0.640pt] (1880.0000,-1740.9336) circle
        (0.9031cm);
      \path[draw=black,line join=miter,line cap=butt,miter limit=4.00,even odd
        rule,line width=0.640pt] (-1680.0001,2192.3621) circle (0.9031cm);
      \path[cm={{-0.23861,-0.97112,0.97112,-0.23861,(0.0,0.0)}},draw=black,line
        join=miter,line cap=butt,miter limit=4.00,even odd rule,line width=0.480pt]
        (-1626.3423,-2161.9302) ellipse (2.0320cm and 0.5080cm);
      \path[cm={{0.63996,-0.7684,0.4865,0.87368,(0.0,0.0)}},draw=black,line
        join=miter,line cap=butt,miter limit=4.00,even odd rule,line width=0.228pt]
        (-2652.9680,118.8193) ellipse (0.7970cm and 0.2911cm);
      \path[cm={{-0.7825,-0.62265,0.87864,-0.47748,(0.0,0.0)}},draw=black,line
        join=miter,line cap=butt,miter limit=4.00,even odd rule,line width=0.257pt]
        (-346.2951,-1979.8077) ellipse (0.9230cm and 0.3201cm);
      \path[draw=black,line join=miter,line cap=butt,miter limit=4.00,even odd
        rule,line width=0.640pt] (-1597.1429,2100.9336) circle (0.9031cm);
      \path[draw=black,line join=miter,line cap=butt,miter limit=4.00,even odd
        rule,line width=0.640pt] (-1402.8572,1189.5050) circle (0.9031cm);
      \path[draw=black,line join=miter,line cap=butt,miter limit=4.00,even odd
        rule,line width=0.640pt] (-1411.4285,2140.9336) circle (0.9031cm);
      \path[cm={{0.89985,0.43621,-0.43621,0.89985,(0.0,0.0)}},draw=black,line
        join=miter,line cap=butt,miter limit=4.00,even odd rule,line width=0.480pt]
        (-419.6118,2573.1099) ellipse (2.0320cm and 0.5080cm);
      \path[scale=-1.000,draw=black,line join=miter,line cap=butt,miter
        limit=4.00,even odd rule,line width=0.640pt] (842.8572,-1720.9336) circle
        (0.9031cm);
      \path[scale=-1.000,draw=black,line join=miter,line cap=butt,miter
        limit=4.00,even odd rule,line width=0.640pt] (1290.5933,-2189.5051) circle
        (0.9031cm);
      \path[cm={{-0.70711,0.70711,-0.70711,-0.70711,(0.0,0.0)}},draw=black,line
        join=miter,line cap=butt,miter limit=4.00,even odd rule,line width=0.480pt]
        (2357.7656,-637.6468) ellipse (2.0320cm and 0.5080cm);
      \path[cm={{-0.97557,0.21967,-0.14792,-0.989,(0.0,0.0)}},draw=black,line
        join=miter,line cap=butt,miter limit=4.00,even odd rule,line width=0.414pt]
        (1349.8927,-1752.2445) ellipse (1.5210cm and 0.5039cm);
      \path[cm={{0.05183,0.99866,-0.98513,0.17183,(0.0,0.0)}},draw=black,line
        join=miter,line cap=butt,miter limit=4.00,even odd rule,line width=0.356pt]
        (1752.8667,1083.2656) ellipse (1.1146cm and 0.5086cm);
      \path[cm={{0.61676,0.78715,-0.86966,0.49365,(0.0,0.0)}},draw=black,line
        join=miter,line cap=butt,miter limit=4.00,even odd rule,line width=0.327pt]
        (1005.9234,1737.4598) ellipse (1.3428cm and 0.3570cm);
      \path[scale=-1.000,draw=black,line join=miter,line cap=butt,miter
        limit=4.00,even odd rule,line width=0.640pt] (1142.0217,-2059.5049) circle
        (0.9031cm);
      \path[scale=-1.000,draw=black,line join=miter,line cap=butt,miter
        limit=4.00,even odd rule,line width=0.640pt] (973.4504,-2009.5050) circle
        (0.9031cm);
      \path[scale=-1.000,draw=black,line join=miter,line cap=butt,miter
        limit=4.00,even odd rule,line width=0.640pt] (979.1648,-1866.6478) circle
        (0.9031cm);
      \path[cm={{-0.70711,0.70711,-0.70711,-0.70711,(0.0,0.0)}},draw=black,line
        join=miter,line cap=butt,miter limit=4.00,even odd rule,line width=0.480pt]
        (1913.2985,-625.5250) ellipse (2.0320cm and 0.5080cm);
      \path[cm={{0.66187,0.74962,-0.79591,0.60542,(0.0,0.0)}},draw=black,line
        join=miter,line cap=butt,miter limit=4.00,even odd rule,line width=0.414pt]
        (505.4849,1673.9463) ellipse (1.5210cm and 0.5039cm);
      \path[cm={{-0.05183,-0.99866,0.98513,-0.17183,(0.0,0.0)}},draw=black,line
        join=miter,line cap=butt,miter limit=4.00,even odd rule,line width=0.356pt]
        (-1348.1738,-1024.2715) ellipse (1.1146cm and 0.5086cm);
      \path[draw=black,line join=miter,line cap=butt,miter limit=4.00,even odd
        rule,line width=0.640pt] (-1073.4504,1332.3621) circle (0.9031cm);
      \path[draw=black,line join=miter,line cap=butt,miter limit=4.00,even odd
        rule,line width=0.640pt] (-942.0220,1449.5048) circle (0.9031cm);
      \path[draw=black,line join=miter,line cap=butt,miter limit=4.00,even odd
        rule,line width=0.640pt] (-936.3076,1592.3621) circle (0.9031cm);
      \path[scale=-1.000,draw=black,line join=miter,line cap=butt,miter
        limit=4.00,even odd rule,line width=0.640pt] (1136.3076,-1140.9335) circle
        (0.9031cm);
      \path[cm={{0.23861,0.97112,-0.97112,0.23861,(0.0,0.0)}},draw=black,line
        join=miter,line cap=butt,miter limit=4.00,even odd rule,line width=0.480pt]
        (938.6842,1368.3796) ellipse (2.0320cm and 0.5080cm);
      \path[cm={{-0.63996,0.7684,-0.4865,-0.87368,(0.0,0.0)}},draw=black,line
        join=miter,line cap=butt,miter limit=4.00,even odd rule,line width=0.228pt]
        (1722.6005,151.9091) ellipse (0.7970cm and 0.2911cm);
      \path[cm={{0.7825,0.62265,-0.87864,0.47748,(0.0,0.0)}},draw=black,line
        join=miter,line cap=butt,miter limit=4.00,even odd rule,line width=0.257pt]
        (1374.1729,2757.6851) ellipse (0.9230cm and 0.3201cm);
      \path[scale=-1.000,draw=black,line join=miter,line cap=butt,miter
        limit=4.00,even odd rule,line width=0.640pt] (1213.4504,-1237.2191) circle
        (0.9031cm);
      \path[cm={{-0.89985,-0.43621,0.43621,-0.89985,(0.0,0.0)}},draw=black,line
        join=miter,line cap=butt,miter limit=4.00,even odd rule,line width=0.480pt]
        (660.6152,-1654.8398) ellipse (2.0320cm and 0.5080cm);
    \path[draw=black,line join=miter,line cap=butt,miter limit=4.00,even odd
      rule,line width=0.640pt] (-1600.0001,1292.3621) circle (0.9031cm);
    \path[cm={{0.64439,-0.7647,0.64439,0.7647,(0.0,0.0)}},draw=black,line
      join=miter,line cap=butt,miter limit=4.00,even odd rule,line width=0.384pt]
      (-1940.8695,-293.8089) ellipse (1.6255cm and 0.4064cm);
    \path[cm={{0.99996,0.00934,-0.09953,0.99503,(0.0,0.0)}},draw=black,line
      join=miter,line cap=butt,miter limit=4.00,even odd rule,line width=0.450pt]
      (-1501.3894,1450.7275) ellipse (1.8025cm and 0.5040cm);
    \path[cm={{-0.45078,-0.89264,0.78843,-0.61513,(0.0,0.0)}},draw=black,line
      join=miter,line cap=butt,miter limit=4.00,even odd rule,line width=0.505pt]
      (-132.7269,-2268.3213) ellipse (1.6103cm and 0.7093cm);
    \path[cm={{-0.62301,-0.78221,0.8731,-0.48755,(0.0,0.0)}},draw=black,line
      join=miter,line cap=butt,miter limit=4.00,even odd rule,line width=0.281pt]
      (-684.8889,-2571.6060) ellipse (1.1489cm and 0.3073cm);
    \path[draw=black,line join=miter,line cap=butt,miter limit=4.00,even odd
      rule,line width=0.640pt] (-1551.4287,1423.7908) circle (0.9031cm);
    \path[draw=black,line join=miter,line cap=butt,miter limit=4.00,even odd
      rule,line width=0.640pt] (-1740.0001,1423.7906) circle (0.9031cm);
    \path[draw=black,line join=miter,line cap=butt,miter limit=4.00,even odd
      rule,line width=0.640pt] (-1685.7145,1578.0764) circle (0.9031cm);
    \path[cm={{0.77315,-0.63423,0.77315,0.63423,(0.0,0.0)}},draw=black,line
      join=miter,line cap=butt,miter limit=4.00,even odd rule,line width=0.665pt]
      (-2458.3303,135.7292) ellipse (2.8155cm and 0.7039cm);
    \path[cm={{-0.97834,-0.207,0.27782,-0.96063,(0.0,0.0)}},draw=black,line
      join=miter,line cap=butt,miter limit=4.00,even odd rule,line width=0.414pt]
      (1119.1946,-2193.2358) ellipse (1.5210cm and 0.5039cm);
    \path[cm={{0.40788,0.91304,-0.85255,0.52264,(0.0,0.0)}},draw=black,line
      join=miter,line cap=butt,miter limit=4.00,even odd rule,line width=0.449pt]
      (719.8888,2420.5334) ellipse (1.4348cm and 0.6297cm);
    \path[scale=-1.000,draw=black,line join=miter,line cap=butt,miter
      limit=4.00,even odd rule,line width=0.640pt] (1625.7144,-1886.6479) circle
      (0.9031cm);
    \path[scale=-1.000,draw=black,line join=miter,line cap=butt,miter
      limit=4.00,even odd rule,line width=0.640pt] (1774.2858,-1843.7908) circle
      (0.9031cm);
    \path[draw=black,line join=miter,line cap=butt,miter limit=4.00,even odd
      rule,line width=0.640pt] (-1700.0002,1726.6478) circle (0.9031cm);
    \path[cm={{0.36273,-0.93189,0.96869,0.24829,(0.0,0.0)}},draw=black,line
      join=miter,line cap=butt,miter limit=4.00,even odd rule,line width=0.430pt]
      (-2356.4368,-722.1372) ellipse (1.5236cm and 0.5442cm);
    \path[cm={{-0.7825,-0.62265,0.87864,-0.47748,(0.0,0.0)}},draw=black,line
      join=miter,line cap=butt,miter limit=4.00,even odd rule,line width=0.257pt]
      (-449.5538,-2156.3132) ellipse (0.9230cm and 0.3201cm);
    \path[draw=black,line join=miter,line cap=butt,miter limit=4.00,even odd
      rule,line width=0.640pt] (-1514.2859,1949.5051) circle (0.9031cm);
    \path[draw=black,line join=miter,line cap=butt,miter limit=4.00,even odd
      rule,line width=0.640pt] (-1477.1429,1338.0764) circle (0.9031cm);
    \path[draw=black,line join=miter,line cap=butt,miter limit=4.00,even odd
      rule,line width=0.640pt] (-1328.5714,1989.5051) circle (0.9031cm);
    \path[cm={{0.89985,0.43621,-0.43621,0.89985,(0.0,0.0)}},draw=black,line
      join=miter,line cap=butt,miter limit=4.00,even odd rule,line width=0.480pt]
      (-411.1078,2400.7046) ellipse (2.0320cm and 0.5080cm);
    \path[scale=-1.000,draw=black,line join=miter,line cap=butt,miter
      limit=4.00,even odd rule,line width=0.640pt] (1022.8572,-1706.6478) circle
      (0.9031cm);
    \path[scale=-1.000,draw=black,line join=miter,line cap=butt,miter
      limit=4.00,even odd rule,line width=0.640pt] (1207.7362,-2038.0767) circle
      (0.9031cm);
    \path[cm={{-0.82498,0.56516,-0.82498,-0.56516,(0.0,0.0)}},draw=black,line
      join=miter,line cap=butt,miter limit=4.00,even odd rule,line width=0.178pt]
      (2131.3291,-301.9961) ellipse (0.7527cm and 0.1882cm);
    \path[cm={{-0.99899,-0.04483,-0.14229,-0.98982,(0.0,0.0)}},draw=black,line
      join=miter,line cap=butt,miter limit=4.00,even odd rule,line width=0.417pt]
      (1346.3007,-1938.1511) ellipse (1.4854cm and 0.5238cm);
    \path[cm={{-0.10234,0.99475,-0.85092,0.5253,(0.0,0.0)}},draw=black,line
      join=miter,line cap=butt,miter limit=4.00,even odd rule,line width=0.497pt]
      (1306.6970,1228.5986) ellipse (1.8865cm and 0.5856cm);
    \path[cm={{0.61676,0.78715,-0.86966,0.49365,(0.0,0.0)}},draw=black,line
      join=miter,line cap=butt,miter limit=4.00,even odd rule,line width=0.327pt]
      (903.5187,1871.8120) ellipse (1.3428cm and 0.3570cm);
    \path[scale=-1.000,draw=black,line join=miter,line cap=butt,miter
      limit=4.00,even odd rule,line width=0.640pt] (1162.0220,-1855.2192) circle
      (0.9031cm);
    \path[cm={{-0.70711,0.70711,-0.70711,-0.70711,(0.0,0.0)}},draw=black,line
      join=miter,line cap=butt,miter limit=4.00,even odd rule,line width=0.480pt]
      (2030.4762,-488.1443) ellipse (2.0320cm and 0.5080cm);
    \path[cm={{0.81106,-0.58497,0.07316,0.99732,(0.0,0.0)}},draw=black,line
      join=miter,line cap=butt,miter limit=4.00,even odd rule,line width=0.205pt]
      (-2205.1895,167.1333) ellipse (0.7222cm and 0.2618cm);
    \path[cm={{-0.62514,-0.78052,0.21132,-0.97742,(0.0,0.0)}},draw=black,line
      join=miter,line cap=butt,miter limit=4.00,even odd rule,line width=0.341pt]
      (1029.5027,-2374.9661) ellipse (0.8986cm and 0.5790cm);
    \path[draw=black,line join=miter,line cap=butt,miter limit=4.00,even odd
      rule,line width=0.640pt] (-1193.4504,1466.6477) circle (0.9031cm);
    \path[draw=black,line join=miter,line cap=butt,miter limit=4.00,even odd
      rule,line width=0.640pt] (-1076.3077,1458.0762) circle (0.9031cm);
    \path[draw=black,line join=miter,line cap=butt,miter limit=4.00,even odd
      rule,line width=0.640pt] (-1116.3076,1578.0763) circle (0.9031cm);
    \path[scale=-1.000,draw=black,line join=miter,line cap=butt,miter
      limit=4.00,even odd rule,line width=0.640pt] (1207.7362,-1298.0763) circle
      (0.9031cm);
    \path[cm={{0.23861,0.97112,-0.97112,0.23861,(0.0,0.0)}},draw=black,line
      join=miter,line cap=butt,miter limit=4.00,even odd rule,line width=0.480pt]
      (1065.2393,1475.9700) ellipse (2.0320cm and 0.5080cm);
    \path[cm={{-0.69738,0.7167,-0.54532,-0.83823,(0.0,0.0)}},draw=black,line
      join=miter,line cap=butt,miter limit=4.00,even odd rule,line width=0.284pt]
      (1832.1473,-36.6036) ellipse (1.0088cm and 0.3581cm);
    \path[cm={{0.7825,0.62265,-0.87864,0.47748,(0.0,0.0)}},draw=black,line
      join=miter,line cap=butt,miter limit=4.00,even odd rule,line width=0.257pt]
      (1272.6327,2572.9548) ellipse (0.9230cm and 0.3201cm);
    \path[scale=-1.000,draw=black,line join=miter,line cap=butt,miter
      limit=4.00,even odd rule,line width=0.640pt] (1293.4504,-1400.0763) circle
      (0.9031cm);
    \path[cm={{-0.89985,-0.43621,0.43621,-0.89985,(0.0,0.0)}},draw=black,line
      join=miter,line cap=butt,miter limit=4.00,even odd rule,line width=0.480pt]
      (662.6527,-1820.9352) ellipse (2.0320cm and 0.5080cm);
  \end{scope}
  \path[fill=black,line join=miter,line cap=butt,line width=0.800pt]
    (320.4565,2514.3599) node[above right] (text4474-6-6) {(d)};
  \path[draw=black,line join=miter,line cap=butt,miter limit=4.00,even odd
    rule,line width=0.640pt] (257.1427,1152.3621) circle (0.9031cm);
  \path[cm={{0.70711,-0.70711,0.70711,0.70711,(0.0,0.0)}},draw=black,line
    join=miter,line cap=butt,miter limit=4.00,even odd rule,line width=0.480pt]
    (-736.0512,994.6502) ellipse (2.0320cm and 0.5080cm);
  \path[cm={{0.97557,-0.21967,0.14792,0.989,(0.0,0.0)}},draw=black,line
    join=miter,line cap=butt,miter limit=4.00,even odd rule,line width=0.414pt]
    (-170.5541,1289.0756) ellipse (1.5210cm and 0.5039cm);
  \path[cm={{-0.05183,-0.99866,0.98513,-0.17183,(0.0,0.0)}},draw=black,line
    join=miter,line cap=butt,miter limit=4.00,even odd rule,line width=0.356pt]
    (-1384.5962,-130.8490) ellipse (1.1146cm and 0.5086cm);
  \path[cm={{-0.61729,-0.78673,0.86996,-0.49312,(0.0,0.0)}},draw=black,line
    join=miter,line cap=butt,miter limit=4.00,even odd rule,line width=0.401pt]
    (-1406.5737,-1170.4763) ellipse (1.6452cm and 0.4376cm);
  \path[draw=black,line join=miter,line cap=butt,miter limit=4.00,even odd
    rule,line width=0.640pt] (108.5713,1292.3622) circle (0.9031cm);
  \path[draw=black,line join=miter,line cap=butt,miter limit=4.00,even odd
    rule,line width=0.640pt] (-60.0001,1332.3621) circle (0.9031cm);
  \path[draw=black,line join=miter,line cap=butt,miter limit=4.00,even odd
    rule,line width=0.640pt] (-200.0001,1620.9335) circle (0.9031cm);
  \path[draw=black,line join=miter,line cap=butt,miter limit=4.00,even odd
    rule,line width=0.640pt] (-54.2858,1475.2192) circle (0.9031cm);
  \path[cm={{0.70711,-0.70711,0.70711,0.70711,(0.0,0.0)}},draw=black,line
    join=miter,line cap=butt,miter limit=4.00,even odd rule,line width=0.480pt]
    (-1180.5183,1006.7721) ellipse (2.0320cm and 0.5080cm);
  \path[cm={{-0.66187,-0.74962,0.79591,-0.60542,(0.0,0.0)}},draw=black,line
    join=miter,line cap=butt,miter limit=4.00,even odd rule,line width=0.414pt]
    (-1534.0876,-1320.6155) ellipse (1.5210cm and 0.5039cm);
  \path[cm={{0.05183,0.99866,-0.98513,0.17183,(0.0,0.0)}},draw=black,line
    join=miter,line cap=butt,miter limit=4.00,even odd rule,line width=0.356pt]
    (1789.2892,189.8433) ellipse (1.1146cm and 0.5086cm);
  \path[scale=-1.000,draw=black,line join=miter,line cap=butt,miter
    limit=4.00,even odd rule,line width=0.640pt] (-39.9999,-2009.5050) circle
    (0.9031cm);
  \path[scale=-1.000,draw=black,line join=miter,line cap=butt,miter
    limit=4.00,even odd rule,line width=0.640pt] (91.4285,-1892.3623) circle
    (0.9031cm);
  \path[scale=-1.000,draw=black,line join=miter,line cap=butt,miter
    limit=4.00,even odd rule,line width=0.640pt] (97.1429,-1749.5050) circle
    (0.9031cm);
  \path[draw=black,line join=miter,line cap=butt,miter limit=4.00,even odd
    rule,line width=0.640pt] (102.8570,2200.9336) circle (0.9031cm);
  \path[cm={{-0.23861,-0.97112,0.97112,-0.23861,(0.0,0.0)}},draw=black,line
    join=miter,line cap=butt,miter limit=4.00,even odd rule,line width=0.480pt]
    (-2060.0684,-432.6140) ellipse (2.0320cm and 0.5080cm);
  \path[cm={{0.63996,-0.7684,0.4865,0.87368,(0.0,0.0)}},draw=black,line
    join=miter,line cap=butt,miter limit=4.00,even odd rule,line width=0.228pt]
    (-987.8532,1593.1045) ellipse (0.7970cm and 0.2911cm);
  \path[cm={{-0.7825,-0.62265,0.87864,-0.47748,(0.0,0.0)}},draw=black,line
    join=miter,line cap=butt,miter limit=4.00,even odd rule,line width=0.257pt]
    (-1279.0530,-781.3988) ellipse (0.9230cm and 0.3201cm);
  \path[draw=black,line join=miter,line cap=butt,miter limit=4.00,even odd
    rule,line width=0.640pt] (185.7142,2109.5051) circle (0.9031cm);
  \path[draw=black,line join=miter,line cap=butt,miter limit=4.00,even odd
    rule,line width=0.640pt] (379.9999,1198.0764) circle (0.9031cm);
  \path[draw=black,line join=miter,line cap=butt,miter limit=4.00,even odd
    rule,line width=0.640pt] (371.4286,2149.5051) circle (0.9031cm);
  \path[cm={{0.89985,0.43621,-0.43621,0.89985,(0.0,0.0)}},draw=black,line
    join=miter,line cap=butt,miter limit=4.00,even odd rule,line width=0.480pt]
    (1188.4227,1803.1241) ellipse (2.0320cm and 0.5080cm);
  \path[scale=-1.000,draw=black,line join=miter,line cap=butt,miter
    limit=4.00,even odd rule,line width=0.640pt] (-939.9999,-1729.5050) circle
    (0.9031cm);
  \path[scale=-1.000,draw=black,line join=miter,line cap=butt,miter
    limit=4.00,even odd rule,line width=0.640pt] (-492.2638,-2198.0767) circle
    (0.9031cm);
  \path[cm={{-0.70711,0.70711,-0.70711,-0.70711,(0.0,0.0)}},draw=black,line
    join=miter,line cap=butt,miter limit=4.00,even odd rule,line width=0.480pt]
    (1103.1562,-1904.3781) ellipse (2.0320cm and 0.5080cm);
  \path[cm={{-0.97557,0.21967,-0.14792,-0.989,(0.0,0.0)}},draw=black,line
    join=miter,line cap=butt,miter limit=4.00,even odd rule,line width=0.414pt]
    (-416.7930,-2153.3213) ellipse (1.5210cm and 0.5039cm);
  \path[cm={{0.05183,0.99866,-0.98513,0.17183,(0.0,0.0)}},draw=black,line
    join=miter,line cap=butt,miter limit=4.00,even odd rule,line width=0.356pt]
    (2069.9702,-709.8260) ellipse (1.1146cm and 0.5086cm);
  \path[cm={{0.61676,0.78715,-0.86966,0.49365,(0.0,0.0)}},draw=black,line
    join=miter,line cap=butt,miter limit=4.00,even odd rule,line width=0.327pt]
    (1903.3331,323.8407) ellipse (1.3428cm and 0.3570cm);
  \path[scale=-1.000,draw=black,line join=miter,line cap=butt,miter
    limit=4.00,even odd rule,line width=0.640pt] (-640.8354,-2068.0764) circle
    (0.9031cm);
  \path[scale=-1.000,draw=black,line join=miter,line cap=butt,miter
    limit=4.00,even odd rule,line width=0.640pt] (-809.4067,-2018.0764) circle
    (0.9031cm);
  \path[scale=-1.000,draw=black,line join=miter,line cap=butt,miter
    limit=4.00,even odd rule,line width=0.640pt] (-803.6923,-1875.2192) circle
    (0.9031cm);
  \path[cm={{-0.70711,0.70711,-0.70711,-0.70711,(0.0,0.0)}},draw=black,line
    join=miter,line cap=butt,miter limit=4.00,even odd rule,line width=0.480pt]
    (658.6890,-1892.2562) ellipse (2.0320cm and 0.5080cm);
  \path[cm={{0.66187,0.74962,-0.79591,0.60542,(0.0,0.0)}},draw=black,line
    join=miter,line cap=butt,miter limit=4.00,even odd rule,line width=0.414pt]
    (1594.5883,339.6053) ellipse (1.5210cm and 0.5039cm);
  \path[cm={{-0.05183,-0.99866,0.98513,-0.17183,(0.0,0.0)}},draw=black,line
    join=miter,line cap=butt,miter limit=4.00,even odd rule,line width=0.356pt]
    (-1665.2773,768.8201) ellipse (1.1146cm and 0.5086cm);
  \path[draw=black,line join=miter,line cap=butt,miter limit=4.00,even odd
    rule,line width=0.640pt] (709.4067,1340.9335) circle (0.9031cm);
  \path[draw=black,line join=miter,line cap=butt,miter limit=4.00,even odd
    rule,line width=0.640pt] (840.8351,1458.0762) circle (0.9031cm);
  \path[draw=black,line join=miter,line cap=butt,miter limit=4.00,even odd
    rule,line width=0.640pt] (846.5495,1600.9335) circle (0.9031cm);
  \path[scale=-1.000,draw=black,line join=miter,line cap=butt,miter
    limit=4.00,even odd rule,line width=0.640pt] (-646.5495,-1149.5049) circle
    (0.9031cm);
  \path[cm={{0.23861,0.97112,-0.97112,0.23861,(0.0,0.0)}},draw=black,line
    join=miter,line cap=butt,miter limit=4.00,even odd rule,line width=0.480pt]
    (1372.4104,-360.9366) ellipse (2.0320cm and 0.5080cm);
  \path[cm={{-0.63996,0.7684,-0.4865,-0.87368,(0.0,0.0)}},draw=black,line
    join=miter,line cap=butt,miter limit=4.00,even odd rule,line width=0.228pt]
    (57.4856,-1322.3760) ellipse (0.7970cm and 0.2911cm);
  \path[cm={{0.7825,0.62265,-0.87864,0.47748,(0.0,0.0)}},draw=black,line
    join=miter,line cap=butt,miter limit=4.00,even odd rule,line width=0.257pt]
    (2306.9307,1559.2761) ellipse (0.9230cm and 0.3201cm);
  \path[scale=-1.000,draw=black,line join=miter,line cap=butt,miter
    limit=4.00,even odd rule,line width=0.640pt] (-569.4067,-1245.7905) circle
    (0.9031cm);
  \path[cm={{-0.89985,-0.43621,0.43621,-0.89985,(0.0,0.0)}},draw=black,line
    join=miter,line cap=butt,miter limit=4.00,even odd rule,line width=0.480pt]
    (-947.4193,-884.8541) ellipse (2.0320cm and 0.5080cm);
  \path[draw=black,line join=miter,line cap=butt,miter limit=4.00,even odd
    rule,line width=0.640pt] (182.8570,1300.9335) circle (0.9031cm);
  \path[cm={{0.64439,-0.7647,0.64439,0.7647,(0.0,0.0)}},draw=black,line
    join=miter,line cap=butt,miter limit=4.00,even odd rule,line width=0.384pt]
    (-563.1017,1095.1678) ellipse (1.6255cm and 0.4064cm);
  \path[cm={{0.99996,0.00934,-0.09953,0.99503,(0.0,0.0)}},draw=black,line
    join=miter,line cap=butt,miter limit=4.00,even odd rule,line width=0.450pt]
    (280.7382,1442.6182) ellipse (1.8025cm and 0.5040cm);
  \path[cm={{-0.45078,-0.89264,0.78843,-0.61513,(0.0,0.0)}},draw=black,line
    join=miter,line cap=butt,miter limit=4.00,even odd rule,line width=0.505pt]
    (-1257.4724,-650.1014) ellipse (1.6103cm and 0.7093cm);
  \path[cm={{-0.62301,-0.78221,0.8731,-0.48755,(0.0,0.0)}},draw=black,line
    join=miter,line cap=butt,miter limit=4.00,even odd rule,line width=0.281pt]
    (-1573.4182,-1163.6356) ellipse (1.1489cm and 0.3073cm);
  \path[draw=black,line join=miter,line cap=butt,miter limit=4.00,even odd
    rule,line width=0.640pt] (231.4284,1432.3622) circle (0.9031cm);
  \path[draw=black,line join=miter,line cap=butt,miter limit=4.00,even odd
    rule,line width=0.640pt] (42.8570,1432.3621) circle (0.9031cm);
  \path[draw=black,line join=miter,line cap=butt,miter limit=4.00,even odd
    rule,line width=0.640pt] (97.1426,1586.6478) circle (0.9031cm);
  \path[cm={{0.77315,-0.63423,0.77315,0.63423,(0.0,0.0)}},draw=black,line
    join=miter,line cap=butt,miter limit=4.00,even odd rule,line width=0.665pt]
    (-1312.1019,1295.4724) ellipse (2.8155cm and 0.7039cm);
  \path[cm={{-0.97834,-0.207,0.27782,-0.96063,(0.0,0.0)}},draw=black,line
    join=miter,line cap=butt,miter limit=4.00,even odd rule,line width=0.414pt]
    (-600.4392,-1831.6062) ellipse (1.5210cm and 0.5039cm);
  \path[cm={{0.40788,0.91304,-0.85255,0.52264,(0.0,0.0)}},draw=black,line
    join=miter,line cap=butt,miter limit=4.00,even odd rule,line width=0.449pt]
    (1666.9658,782.4335) ellipse (1.4348cm and 0.6297cm);
  \path[scale=-1.000,draw=black,line join=miter,line cap=butt,miter
    limit=4.00,even odd rule,line width=0.640pt] (-157.1428,-1895.2194) circle
    (0.9031cm);
  \path[scale=-1.000,draw=black,line join=miter,line cap=butt,miter
    limit=4.00,even odd rule,line width=0.640pt] (-8.5713,-1852.3622) circle
    (0.9031cm);
  \path[draw=black,line join=miter,line cap=butt,miter limit=4.00,even odd
    rule,line width=0.640pt] (82.8569,1735.2192) circle (0.9031cm);
  \path[cm={{0.36273,-0.93189,0.96869,0.24829,(0.0,0.0)}},draw=black,line
    join=miter,line cap=butt,miter limit=4.00,even odd rule,line width=0.430pt]
    (-1918.9171,954.5186) ellipse (1.5236cm and 0.5442cm);
  \path[cm={{-0.7825,-0.62265,0.87864,-0.47748,(0.0,0.0)}},draw=black,line
    join=miter,line cap=butt,miter limit=4.00,even odd rule,line width=0.257pt]
    (-1382.3116,-957.9043) ellipse (0.9230cm and 0.3201cm);
  \path[draw=black,line join=miter,line cap=butt,miter limit=4.00,even odd
    rule,line width=0.640pt] (268.5712,1958.0765) circle (0.9031cm);
  \path[draw=black,line join=miter,line cap=butt,miter limit=4.00,even odd
    rule,line width=0.640pt] (305.7141,1346.6478) circle (0.9031cm);
  \path[draw=black,line join=miter,line cap=butt,miter limit=4.00,even odd
    rule,line width=0.640pt] (454.2857,1998.0765) circle (0.9031cm);
  \path[cm={{0.89985,0.43621,-0.43621,0.89985,(0.0,0.0)}},draw=black,line
    join=miter,line cap=butt,miter limit=4.00,even odd rule,line width=0.480pt]
    (1196.9268,1630.7189) ellipse (2.0320cm and 0.5080cm);
  \path[scale=-1.000,draw=black,line join=miter,line cap=butt,miter
    limit=4.00,even odd rule,line width=0.640pt] (-759.9999,-1715.2192) circle
    (0.9031cm);
  \path[scale=-1.000,draw=black,line join=miter,line cap=butt,miter
    limit=4.00,even odd rule,line width=0.640pt] (-575.1209,-2046.6481) circle
    (0.9031cm);
  \path[cm={{-0.82498,0.56516,-0.82498,-0.56516,(0.0,0.0)}},draw=black,line
    join=miter,line cap=butt,miter limit=4.00,even odd rule,line width=0.178pt]
    (1058.3679,-1390.1237) ellipse (0.7527cm and 0.1882cm);
  \path[cm={{-0.99899,-0.04483,-0.14229,-0.98982,(0.0,0.0)}},draw=black,line
    join=miter,line cap=butt,miter limit=4.00,even odd rule,line width=0.417pt]
    (-448.6964,-1865.5155) ellipse (1.4854cm and 0.5238cm);
  \path[cm={{-0.10234,0.99475,-0.85092,0.5253,(0.0,0.0)}},draw=black,line
    join=miter,line cap=butt,miter limit=4.00,even odd rule,line width=0.497pt]
    (2497.3652,-1009.8248) ellipse (1.8865cm and 0.5856cm);
  \path[cm={{0.61676,0.78715,-0.86966,0.49365,(0.0,0.0)}},draw=black,line
    join=miter,line cap=butt,miter limit=4.00,even odd rule,line width=0.327pt]
    (1800.9285,458.1929) ellipse (1.3428cm and 0.3570cm);
  \path[scale=-1.000,draw=black,line join=miter,line cap=butt,miter
    limit=4.00,even odd rule,line width=0.640pt] (-620.8351,-1863.7906) circle
    (0.9031cm);
  \path[cm={{-0.70711,0.70711,-0.70711,-0.70711,(0.0,0.0)}},draw=black,line
    join=miter,line cap=butt,miter limit=4.00,even odd rule,line width=0.480pt]
    (775.8668,-1754.8755) ellipse (2.0320cm and 0.5080cm);
  \path[cm={{0.81106,-0.58497,0.07316,0.99732,(0.0,0.0)}},draw=black,line
    join=miter,line cap=butt,miter limit=4.00,even odd rule,line width=0.205pt]
    (-118.1949,1399.8302) ellipse (0.7222cm and 0.2618cm);
  \path[cm={{-0.62514,-0.78052,0.21132,-0.97742,(0.0,0.0)}},draw=black,line
    join=miter,line cap=butt,miter limit=4.00,even odd rule,line width=0.341pt]
    (-1218.5729,-588.5378) ellipse (0.8986cm and 0.5790cm);
  \path[draw=black,line join=miter,line cap=butt,miter limit=4.00,even odd
    rule,line width=0.640pt] (589.4067,1475.2191) circle (0.9031cm);
  \path[draw=black,line join=miter,line cap=butt,miter limit=4.00,even odd
    rule,line width=0.640pt] (706.5494,1466.6476) circle (0.9031cm);
  \path[draw=black,line join=miter,line cap=butt,miter limit=4.00,even odd
    rule,line width=0.640pt] (666.5495,1586.6477) circle (0.9031cm);
  \path[scale=-1.000,draw=black,line join=miter,line cap=butt,miter
    limit=4.00,even odd rule,line width=0.640pt] (-575.1209,-1306.6477) circle
    (0.9031cm);
  \path[cm={{0.23861,0.97112,-0.97112,0.23861,(0.0,0.0)}},draw=black,line
    join=miter,line cap=butt,miter limit=4.00,even odd rule,line width=0.480pt]
    (1498.9655,-253.3462) ellipse (2.0320cm and 0.5080cm);
  \path[cm={{-0.69738,0.7167,-0.54532,-0.83823,(0.0,0.0)}},draw=black,line
    join=miter,line cap=butt,miter limit=4.00,even odd rule,line width=0.284pt]
    (304.8047,-1352.7469) ellipse (1.0088cm and 0.3581cm);
  \path[cm={{0.7825,0.62265,-0.87864,0.47748,(0.0,0.0)}},draw=black,line
    join=miter,line cap=butt,miter limit=4.00,even odd rule,line width=0.257pt]
    (2205.3906,1374.5459) ellipse (0.9230cm and 0.3201cm);
  \path[scale=-1.000,draw=black,line join=miter,line cap=butt,miter
    limit=4.00,even odd rule,line width=0.640pt] (-489.4067,-1408.6477) circle
    (0.9031cm);
  \path[cm={{-0.89985,-0.43621,0.43621,-0.89985,(0.0,0.0)}},draw=black,line
    join=miter,line cap=butt,miter limit=4.00,even odd rule,line width=0.480pt]
    (-945.3818,-1050.9495) ellipse (2.0320cm and 0.5080cm);
  \path[cm={{0.6985,-0.71561,0.66274,0.74885,(0.0,0.0)}},draw=black,line
    join=miter,line cap=butt,miter limit=4.00,even odd rule,line width=0.414pt]
    (-1008.1724,1288.9022) ellipse (1.5210cm and 0.5039cm);
  \path[cm={{-0.02942,-0.99957,0.9558,-0.29401,(0.0,0.0)}},draw=black,line
    join=miter,line cap=butt,miter limit=4.00,even odd rule,line width=0.477pt]
    (-1580.7015,181.5261) ellipse (1.9560cm and 0.5220cm);
  \path[cm={{-0.61729,-0.78673,0.86996,-0.49312,(0.0,0.0)}},draw=black,line
    join=miter,line cap=butt,miter limit=4.00,even odd rule,line width=0.401pt]
    (-1621.8063,-850.2687) ellipse (1.6452cm and 0.4376cm);
  \path[draw=black,line join=miter,line cap=butt,miter limit=4.00,even odd
    rule,line width=0.640pt] (211.4286,1632.3621) circle (0.9031cm);
  \path[cm={{0.70711,-0.70711,0.70711,0.70711,(0.0,0.0)}},draw=black,line
    join=miter,line cap=butt,miter limit=4.00,even odd rule,line width=0.480pt]
    (-897.6757,1305.7772) ellipse (2.0320cm and 0.5080cm);
  \path[cm={{-0.66187,-0.74962,0.79591,-0.60542,(0.0,0.0)}},draw=black,line
    join=miter,line cap=butt,miter limit=4.00,even odd rule,line width=0.414pt]
    (-1517.7549,-1109.5942) ellipse (1.5210cm and 0.5039cm);
  \path[cm={{-0.36268,0.93191,-0.99339,-0.11482,(0.0,0.0)}},draw=black,line
    join=miter,line cap=butt,miter limit=4.00,even odd rule,line width=0.467pt]
    (1873.7484,-974.5889) ellipse (1.8758cm and 0.5204cm);
  \path[scale=-1.000,draw=black,line join=miter,line cap=butt,miter
    limit=4.00,even odd rule,line width=0.640pt] (-605.7142,-1700.9336) circle
    (0.9031cm);
  \path[scale=-1.000,draw=black,line join=miter,line cap=butt,miter
    limit=4.00,even odd rule,line width=0.640pt] (-314.2857,-1760.9336) circle
    (0.9031cm);
  \path[draw=black,line join=miter,line cap=butt,miter limit=4.00,even odd
    rule,line width=0.640pt] (368.5712,1509.5050) circle (0.9031cm);
  \path[cm={{-0.28187,-0.95945,0.97953,-0.20129,(0.0,0.0)}},draw=black,line
    join=miter,line cap=butt,miter limit=4.00,even odd rule,line width=0.440pt]
    (-1469.8029,-81.6766) ellipse (1.7219cm and 0.5042cm);
  \path[cm={{-0.85486,-0.51886,0.34346,-0.93917,(0.0,0.0)}},draw=black,line
    join=miter,line cap=butt,miter limit=4.00,even odd rule,line width=0.505pt]
    (-964.6484,-1103.2455) ellipse (1.6103cm and 0.7093cm);
  \path[cm={{-0.62301,-0.78221,0.8731,-0.48755,(0.0,0.0)}},draw=black,line
    join=miter,line cap=butt,miter limit=4.00,even odd rule,line width=0.281pt]
    (-1786.8259,-844.6866) ellipse (1.1489cm and 0.3073cm);
  \path[draw=black,line join=miter,line cap=butt,miter limit=4.00,even odd
    rule,line width=0.640pt] (508.5711,1598.0764) circle (0.9031cm);
  \path[cm={{0.71569,-0.69842,0.94845,-0.31694,(0.0,0.0)}},draw=black,line
    join=miter,line cap=butt,miter limit=4.00,even odd rule,line width=0.323pt]
    (-3890.0752,3417.4167) ellipse (1.2456cm and 0.3752cm);
  \path[cm={{-0.99861,0.05272,0.4415,-0.89726,(0.0,0.0)}},draw=black,line
    join=miter,line cap=butt,miter limit=4.00,even odd rule,line width=0.478pt]
    (-1478.9811,-2157.7256) ellipse (1.7881cm and 0.5715cm);
  \path[cm={{0.9998,-0.02003,0.17767,0.98409,(0.0,0.0)}},draw=black,line
    join=miter,line cap=butt,miter limit=4.00,even odd rule,line width=0.486pt]
    (-321.6482,1769.7924) ellipse (1.6754cm and 0.6317cm);
  \path[scale=-1.000,draw=black,line join=miter,line cap=butt,miter
    limit=4.00,even odd rule,line width=0.640pt] (-419.9998,-1863.7908) circle
    (0.9031cm);
  \path[draw=black,line join=miter,line cap=butt,miter limit=4.00,even odd
    rule,line width=0.640pt] (494.2854,1746.6478) circle (0.9031cm);
  \path[scale=-1.000,draw=black,line join=miter,line cap=butt,miter
    limit=4.00,even odd rule,line width=0.640pt] (-405.7143,-1689.5050) circle
    (0.9031cm);
  \path[cm={{0.5989,0.80082,-0.85949,0.51115,(0.0,0.0)}},draw=black,line
    join=miter,line cap=butt,miter limit=4.00,even odd rule,line width=0.319pt]
    (1705.8148,537.7660) ellipse (1.3200cm and 0.3448cm);
  \path[cm={{-0.63701,0.77085,-0.63701,-0.77085,(0.0,0.0)}},draw=black,line
    join=miter,line cap=butt,miter limit=4.00,even odd rule,line width=0.345pt]
    (811.6473,-1528.3491) ellipse (1.4612cm and 0.3653cm);
  \path[cm={{-0.04019,-0.99919,0.9756,-0.21956,(0.0,0.0)}},draw=black,line
    join=miter,line cap=butt,miter limit=4.00,even odd rule,line width=0.405pt]
    (-1900.9943,546.3368) ellipse (1.4350cm and 0.5127cm);
  \path[cm={{0.9807,-0.19552,0.04843,0.99883,(0.0,0.0)}},draw=black,line
    join=miter,line cap=butt,miter limit=4.00,even odd rule,line width=0.327pt]
    (601.0002,1829.1567) ellipse (1.3428cm and 0.3570cm);
  \path[cm={{-0.70711,0.70711,-0.70711,-0.70711,(0.0,0.0)}},draw=black,line
    join=miter,line cap=butt,miter limit=4.00,even odd rule,line width=0.480pt]
    (1121.3389,-1449.8094) ellipse (2.0320cm and 0.5080cm);
  \path[cm={{-0.03438,-0.99941,0.96709,-0.25445,(0.0,0.0)}},draw=black,line
    join=miter,line cap=butt,miter limit=4.00,even odd rule,line width=0.440pt]
    (-1621.1812,463.1957) ellipse (1.6754cm and 0.5166cm);
  \path[cm={{-0.99722,-0.07447,0.04778,-0.99886,(0.0,0.0)}},draw=black,line
    join=miter,line cap=butt,miter limit=4.00,even odd rule,line width=0.430pt]
    (-928.0042,-1650.8502) ellipse (1.6307cm and 0.5085cm);
  \path[cm={{0.89599,0.44408,-0.97656,0.21526,(0.0,0.0)}},draw=black,line
    join=miter,line cap=butt,miter limit=4.00,even odd rule,line width=0.559pt]
    (2688.7085,2155.9055) ellipse (2.0170cm and 0.6937cm);
  \path[cm={{0.68114,0.73215,-0.80612,0.59176,(0.0,0.0)}},draw=black,line
    join=miter,line cap=butt,miter limit=4.00,even odd rule,line width=0.213pt]
    (1659.5432,850.0729) ellipse (0.7858cm and 0.2586cm);

\end{tikzpicture}

\caption{(a) A shell  is collapsing in empty space; in its  classical evolution it would create a horizon when it reached the dotted circle.  (b)  In the theory with fuzzballs, there is a nucleation of `bubbles' as the shell comes close to this dotted circle. Since the shell loses some  energy  in creating these bubbles, the location where the classical horizon would form moves to a smaller radius. (c) The shell keeps moving inwards, losing more and more energy to nucleated bubbles, and thus always staying outside its horizon. (d) We finally get a fuzzball with no horizon or singularity.} \label{fig2qq}

\end{figure}

A similar effect is postulated for the behavior of an infalling shell of energy $E$ in the fuzzball spacetime created by $F_i(M)$. At large $r$, the shell travels as if the spacetime had an infinite depth, and this behavior reproduces infall in the semiclassical geometry. At some radius $r>r_f$ the amplitude of the fields describing the shell becomes comparable to the depth $d$ of spacetime, and here the shell is forced to depart from its semiclassical evolution and transition to a linear combination of fuzzball states $F_j(M+E)$.  In \cite{mat-causality} it was argued that causality can be preserved only if this transition happens at $r\ge 2(M+E)$; we will assume that it happens at the closest point allowed by causality: 
\be
r=2(M+E)
\ee
 Thus the virtual fluctuations of fuzzballs create a `depth' for the spacetime such that pulses with different energy travel differently on the spacetime; at any location $r>r_f$ pulses with sufficiently small energy see semiclassical spacetime, while pulses with energy large enough to form a horizon outside radius $r$ feel the finite depth of the sea and transition to fuzzballs (fig.\ref{fig3qq}). Thus in no case do we develop a horizon, while causality is maintained at all times in the theory. 
 
\b

(F5): In flat Minkowki spacetime we have virtual fuzzball fluctuations of arbitrarily large mass $M$ around each point $x$. But we can consider wavefunctionals where the fluctuations are limited to a mass
\be
M<M_{max}
\label{bbfour}
\ee
Then the virtual fuzzballs have a radius
\be
r<r_{max}=2M_{max}
\label{bbninet}
\ee

\section{The wavefunctional for case C2}\label{secsix}

Let us first consider the case C2 listed in section \ref{secc}. We start with a large, homogeneous, low density dust ball in asymptotically flat spacetime, and let it collapse. In classical general relativity, the interior of this dust ball gives the uniform dust cosmology with the arrow of time reversed: $t\r -t$. We assume that the initial velocities of the dust grains have been tuned so that the dust cosmology is the flat one (\ref{one}). 

The overall picture will be as follows.  By the property (F4) of fuzzballs (listed in section \ref{secp}) the collapsing dust ball does not shrink to a radius $R<r_h=2GM$; it tunnels to fuzzballs of radius $R\approx r_h$ as the $R$ starts to approach the horizon radius $r_h$. Thus there is never a situation where a ball is strictly contained within its horizon radius.

Now consider the Hilbert space $H_F$ of all fuzzball states; this is a very large space, as the number of fuzzballs is given by $Exp[S_{bek}(M)]$ where $S_{bek}$ is the Bekenstein entropy. The dynamical evolution of the matter inside the horizon will be an evolution on this large space $H_F$. The conjecture of fuzzball complementarity says that the low energy dynamics on $H_F$ reproduces, {\it to a good approximation}, the low energy dynamics of string theory around the {\it vacuum}. This conjecture therefore allows an approximate picture where a particle falling onto the fuzzball excites complicated degrees of freedom, but the evolution of these degrees of freedom approximately mimics infall into the classical metric of the black hole.

Let us now list these notions in more detail:

\b

(a) The flat cosmology (\ref{one}) has an infinite spatial slice. But we have regularized this by taking  a ball of proper radius $R_i$;  at the end we will take $R_i$ to infinity. Let the density of the ball be $\rho_0$, and take it to be collapsing with a rate that would be achieved if this ball started from rest when its radius was infinity. Outside this ball we will take empty spacetime. Thus for $R>R_i$ we have the Schwarzschild solution with a mass $M$, where
\be
M=\rho_0{1\over d}\Omega_{d-1}R_i^{d}
\ee

\bigskip

(b) The horizon radius for this mass is, using (\ref{qthir}) 
\be
R_h^{d-2}={16\pi GM_i\over (d-1)\Omega_{d-1}}
\ee
We assume that our starting radius satisfies
\be
R_i>R_h
\ee

\bigskip

(c) When the ball collapses to the radius $R=R_i$, then it will tunnel into fuzzballs, by property (F4). Its further collapse under the semiclassical description is no longer valid. But by the conjecture of fuzzball complementarity evolution in the space of fuzzballs $H_F$ gives an approximate dual description of continued collapse of the ball. Let the time at this point be 
\be
t=t_i
\ee
(Note that $t_i<0$ since we have defined $t=0$ to be the time of the big crunch.)

\b

(d) We can focus on a small subset of the initial ball; for concreteness we let the subset be a ball which has radius at time $T_i$ a radius
\be
R_1\ll R_h
\ee
The mass of the ball is
\be
M_1=\rho_0(t_i) {1\over d}\Omega_{d-1} R_1^d
\ee
We can also write this as
\be
M_1=\left ( {R_1\over R_h}\right )^d
\ee
In the approximate complementary description, the evolution of this ball in the subspace $H_F$ will map to ordinary low energy dynamics of the dust particles in the ball. But note that these particles in the ball are not real particles in the full quantum theory; they are approximate descriptions of waves in $H_F$.

\bigskip

(e) We follow the evolution of this ball until it becomes so dense that it would itself form a black hole.  This happens when the proper radius of the ball is
\be
R_2=R_h\left ( {M_1\over M_i}\right ) ^{1\over d-2}
\ee
At this point the `simulation' of quantum gravity that we had in the space $H_F$ would tell us that the dust ball tunnels into fuzzballs. Let this space of fuzzballs be called $H_{F_1}$. We can follow this fuzzball description of $H_{F_1}$, or we can focus on a small subset
with proper radius
\be
R_3 \ll R_2
\ee
This subset of radius $R_3$ would look like ordinary dust in the approximation where we look at evolution in $H_{F_1}$ in a dual description. 

\bigskip

\begin{figure}
\begin{center}
 \includegraphics[scale=.6] {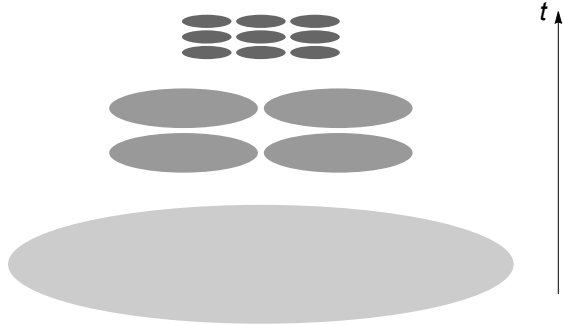}
\end{center}
\caption{A collapsing dust ball stays at its horizon radius, but approximate effective semiclassical infall can be obtained (through fuzzball complementarity) for  small patches. When these smaller patches reach their horizon radius,  we can get an  approximate complementary description of an even smaller patch etc.} 
\label{f5}
\end{figure}

(f) We can continue in this fashion, focusing on smaller and smaller subsets of our initial dust ball, and getting an effective dual description of them in some space $H_{F_k}$. Note that the `real' exact string theory description is only in the original space $H_F$. We then have a subset of this space, which contains approximate fuzzballs described a space $H_{F_1}$. A subset of these approximate degrees of freedom can be described by an approximate space of fuzzballs $H_{F_2}$, and so on (fig.\ref{f5})
\be
H_F \supset H_{F_1} \supset H_{F_2} \supset \dots 
\ee

\section{The cosmological constant}\label{secseven}

In case C2 the spacetime at infinity is the  asymptotically flat vacuum. Thus for the cosmological constant we have the value $\Lambda=0$. Note that by property (F5) we have virtual fuzzball fluctuations of arbitrarily large size at spatial infinity. 

Now consider case C1. This time we are looking for a homogeneous infinite cosmology, and so spacetime is not asymptotically flat. Take a ball of radius $R$ with mass $M$. For sufficiently large $R$, we will have $R<2GM$, so we will have compressed matter to a size where it is inside its horizon. The fuzzball paradigm had said that the information paradox is avoided because we can never compress matter such that it lies within its own horizon. How do we avoid this conflict between our desire to get a homogeneous infinite cosmology and our need to escape the information paradox?

The key is equation (\ref{bbfour}) of property (F5), which says that we can choose a different kind of wavefunctional: one where the virtual fuzzball fluctuations exist only for $M<M_{max}$ rather than for arbitrarily large $M$. In semi-classical gravity we would not consider such virtual fuzzballs, and we would therefore miss this possibility of having different kinds of wavefunctionals for the asymptotically flat black hole problem (where $M_{max}=\infty$) and the homogeneous cosmology (where $M_{max}$ is finite). 

We will now see that taking wavefunctionals with finite $M_{max}$ in (\ref{bbfour}) help resolve several puzzles in quantum gravity.

\subsection{The cosmological constant}

\begin{figure}[h]
\begin{center}
 \includegraphics[scale=.2] {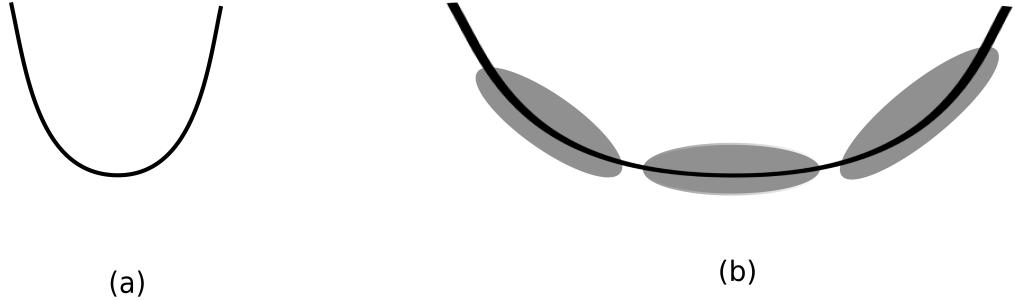}
\end{center}
\caption{ (a) We assume that the quantum fluctuations of local fields generates a positive cosmological constant  which by itself would curve the universe to a small curvature radius (b) The full quantum gravity wavefunctional $\Psi$ also contains virtual fluctuations of extended incompressible objects (virtual fuzzballs, shown as grey blobs) which resist this curving of space, and thus force spacetime to have a large curvature radius. The virtual fuzzball component of $\Psi$ is important because of the large phase space of the $Exp[S_{bek}(M)]$ for fuzzballs of mass $M$.} 
\label{figc}
\end{figure}

A fundamental problem in quantum gravity is the cosmological constant $\Lambda$. The quantum fluctuations of a quantum field give an energy $\h \hbar \omega$ for each mode with frequency $\omega$. If we allow $\omega$ to take arbitrarily large values, then we get  a divergent energy density, and thus $\Lambda=\infty$. If we cut off $\omega$ at the planck scale $m_p$, then we get $\Lambda \sim m_p^4$, which would cause the universe to curl up with a  curvature length scale $l_p$. 
We can try to renormalize away the divergent sum over $\omega$, getting a logarithmic divergence 
$\Lambda \sim m^4\log (m^2/\mu^2)$ for a scalar field of mass $m$ and assuming a renormalization scale $\mu$ (see \cite{martin} for a review). Standard model parameters suggest  that $\lambda$ is of order $(GeV)^4$. Again, this would give a high curvature to the universe, in contrast to the almost flat spacetime we see around us. What cancels the large contribution to $\Lambda$ from these quantum fluctuations? We can of course say that there is a `bare $\Lambda$' which cancels the large contribution of quantum fluctuations, but a close cancellation would need fine tuning, and therefore would not have a clear motivation. 

We will now propose that a different effect flattens out spacetime: an effect arising from the existence of virtual fuzzballs. First consider the quantum fluctuations of usual quantum fields, and suppose that these contribute a large positive energy density $\rho$, leading to a large positive $\Lambda$. This would then tend to generate de-Sitter spacetime with a small horizon radius
\be
r_{dS}= (G\Lambda)^{-\h}
\ee
Now suppose we require that our wavefunctional have virtual fuzzball fluctuations only upto a mass $M_{max}$. Setting $R=2GM$ as the rough radius of a fuzzball of mass $M$, we can write this limit as
\be
R<R_{max}
\ee
where
\be
R_{max}=2GM_{max}
\ee
Thus we have virtual fuzzballs for radius $R<R_{max}$. 

Now we come to the crucial issue. If we have chosen $R_{max}\ll r_{dS}$ then we have no difficulty in constructing a wavefunctional with such virtual fuzzball fluctuations. But we cannot have $R_{max}\gtrsim r_{dS}$. That is, fuzzballs with radius larger than the de Sitter horizon cannot `fit' in the spacetime. We can see this by looking at the Schwarzschild - de Sitter metric:
\be
ds^2=-f(r) dt^2+{dr^2\over f(r)}+r^2 d\Omega^2
\label{bbsix}
\ee
with
\be
f(r)=1-{2GM\over r} -G\Lambda r^2
\label{bbseven}
\ee
Virtual fuzzballs of mass $M$ must satisfy the Gauss constraint generating the exterior Schwarzschild de Sitter geometry just as a real spherically symmetric object of mass $M$ would. Thus we can let $M$ in (\ref{bbseven}) be the mass of the fuzzball. The vanishing of $f$ gives two horizons: the smaller value of $r$ gives the  black hole horizon and the larger value gives the cosmological horizon. The largest allowed value of $M$  occurs when these two horizons have the same value if $r$, which happens for
\be
r=({1\over 3G\Lambda})^{\h}
\label{bbthir}
\ee
Thus we cannot have $R_{max}>({1\over 3G\Lambda})^\h$. Conversely, if we do require a given value of $R_{max}$, then we must have
\be
\Lambda < {1\over 3GR_{max}^2}
\label{bbten}
\ee
We can interpret this result physically as follows. The local quantum fluctuations of quantum fields give rise to a large contribution to $\Lambda$, which we assume to be positive. This tends to curl up the space with a small  curvature length scale. But the part of the gravity wavefunctional $\Psi$ which resides in virtual fuzzballs resists this curling up. By property (F2) the virtual fuzzballs are `compression resistant'. The curled up de Sitter geometry has a structure where the set of points at radial distance $r$  have an area smaller than they would have in flat space. Thus the curled up space cannot tends to compress fuzzballs with a radius larger than the curvature length scale. The virtual fuzzballs resist this compression, and thereby prevent the curling up of the space to a radius smaller than $R_{max}$.

In short, we have proposed a mechanism to get a small effective $\Lambda$ that is different from the usual mechanisms suggested for this purpose. In the traditional computations, we look only at the quantum fluctuations of local fields, and find a large value; we assume that this contribution is positive. We might then try to cancel this large value by a bare value that is fine tuned to be appropriately negative. In our proposal however, there is another source of quantum fluctuations that must be considered: the fluctuations of extended objects that are compression resistant; i.e., fuzzballs. If the quantum wavefunctional $\Psi$ contains these fuzzballs upto radius $R_{max}$, the the space cannot curl to a radius less that $\sim R_{max}$. Thus these virtual fuzzballs flatten space out, cancelling the curling tendency of the local quantum fluctuations. 

In what follows we will assume that in the absence of other matter, spacetime does curl up to the maximum extent possible that is allowed by the virtual fuzzballs; thus from (\ref{bbten}) we get
\be
\Lambda ={1\over 3GR_{max}^2}
\label{bbel}
\ee

\subsection{Prevention of the `bags of gold' problem}

Consider again the formation of a black hole in asymptotically flat spacetime. There are virtual fuzzball fluctuations of arbitrarily large radius $R$. Horizon formation is avoided because if a horizon did form, then by causality the virtual fuzzballs inside the horizon would have to get monotonically pushed to smaller and smaller radii, and this is not allowed since they are `compression resistant'. Thus a pulse of energy collapsing to make the hole feels a departure from semiclassical evolution as it approaches horizon formation, due to the virtual fuzzballs in the wavefunctional resisting compression. This is the effect that we called `hitting the bottom of the sea' when we modeled spacetime as having a `thickness'. This effect removes the information paradox if we start with no black hole: the hole is prevented from forming and fuzzballs form instead. Note that causality is maintained throughout; we never require any physical effects to travel outside the light cone. 

But now consider a different way of approaching the black hole paradox. Assume usual semiclassical physics. We {\it start} with a ball of mass $M$ and radius $r_b<2GM$. The ball is inside its horizon $r_h=2GM$. In the region $r_b<r<r_h$ the light cones point inwards, and if we assume that causality holds to leading order, then the matter at $r<r_b$ cannot emerge to the band  $r_b<r<r_h$, and the light cone structure in the band cannot change to leading order. Then entangled pairs will be created at the horizon, and by the small correction theorem (\ref{weightq}) , we will not be able to escape the entanglement problem that is central to the information paradox. (The geometry created by a ball wth size $r_b<2GM$ is sometimes called a `bag of gold' as it has the structure of  mass at the end of a long throat-type region.)

Let us see how this version of the information paradox is resolved by our postulate of the virtual fuzzball component of the wavefunctional $\Psi$. We consider two possibilities:

\b

(i) The wavefunctional $\Psi$ has virtual fuzzballs of radii 
\be
R\gtrsim r_h
\ee
In this case there  will be fuzzballs with radii 
\be
r_b<R<r_h
\label{bbeight}
\ee
in the band $r_b<r<r_h$. By causality, these fuzzballs will be forced to evolve to configurations with smaller radius, which is disallowed since the fuzzballs are resist compression. Thus the wavefunctional will evolve away almost immediately from the vacuum configuration in the band  (\ref{bbeight}), and we will not be able to argue for the usual production of entangled pairs which are central to the Hawking process. This removes the paradox.

\b

(ii) The wavefunctional $\Psi$ has virtual fuzzballs only with radii
\be
R\ll  r_h
\label{bbnine}
\ee
Consider the wavefunctional $\Psi$ in the region $r<r_h$. With (\ref{bbnine}), we get, using (\ref{bbel})
\be
\Lambda>{1\over 3GR^2}\gg {1\over 3 G r_h^2}
\ee
which can be rewritten as
\be
r_h\gg ({1\over 3G\Lambda})^{\h}
\label{bbfourt}
\ee
Note that we are seeking to place a band of the Schwarzschild geometry with horizon radius $r_h$ in the band (\ref{bbeight}). But from the Schwarzschild-de Sitter geometry (\ref{bbsix}) and (\ref{bbthir}) we see that the  radius $r_h$ for such a geometry is limited as
\be
r_h<({1\over 3G\Lambda})^{\h}
\label{bbfift}
\ee
We see that (\ref{bbfourt}) and (\ref{bbfift}) are in contradiction with each other. Therefore we conclude that we cannot have a situation where the ball of mass $M$ is compressed to a radius $r_b<r_h=2GM$, even if we allow the wavefunctional $\Psi$ in the vacuum regions to have a bounded range for the radii $R$ of virtual fuzzballs. 

\b

Let us summarize the arguments in (i) and (ii) above in physical terms. In the semiclassical picture, if we compress the mass $M$ to a radius $r_b<r_h=2GM$, then we create a band of empty space $r_b<r<r_h$. In our postulate about the nature of the quantum gravity wavefunctional $\Psi$, we have encountered a new component: virtual fuzzballs, which are extended, compression resistant objects. We have to now ask what is the nature of this component of $\Psi$ in the band $r_b<r<r_h$. In possibility (i) we allow virtual fuzzballs with radius $R\gtrsim r_h$. But such fuzzballs cannot be compressed into a region with radius $r\lesssim r_h$, for the following reason. We have assumed that causality holds, so the light cones in the band point inwards, and any virtual fuzzballs structure in the band is forced to compress to ever smaller radii. Since this compression is disallowed by the structure of the fuzzball, we cannot have a vacuum region in the band. We can try to evade this difficulty be taking possibility (ii) where we have virtual fuzzballs only with small radius $R\ll r_h$. These small fuzzballs do not have to be centered around $r=0$, and can avoid getting compressed, at least for some part of their evolution in the interior of the black hole geometry. But since the band is a vacuum region, the absence of fuzzballs of radius $\sim r_h$ and larger forces the spacetime to curl up  with a curvature radius smaller than $r_h$. In this curled up de Sitter space, we cannot fit a band with the Schwarzschild geometry of a black hole with radius $r_h$. So we again conclude that we cannot make a `bag of gold' configuration where the mass $M$ is compressed to a radius $r_b<r_h=2GM$ with vacuum in the band $r_b<r<r_h$.

\section{The wavefunctional for the case C1}\label{seccone}

Let us now put together what we have learnt about wavefunctionals with limited virtual fuzzball size (\ref{bbninet}) to get a picture of the wavefunctional $\Psi$ for an infinite cosmology:

\b

(a) We have a flat infinite cosmology of the form (\ref{one}). Let the density of matter be $\rho_m$. The horizon radius is given by 
\be
R_h=H^{-1}=\left ( {8\pi G\over 3}\rho_m\right )^{-\h}
\ee
This is the information from  semiclassical gravity. We now have to address the virtual fuzzball component of the full quantum wavefunctional $\Psi$. We conjecture that there cannot be virtual fuzzballs of size larger than $R_h$. It {\it is} possible to set the maximal virtual fuzzball radius $R_{max}$ to be smaller than $R_h$, but to start we will set
\be
R_{max}=R_h
\label{bbtone}
\ee
In physical terms, we are saying the following. In flat empty spacetime we had conjectured that there are virtual fuzzballs with  arbitrarily large radius $R$.  These virtual fuzzballs ensured that we get Minkowski spacetime as the semiclassical description of our state. If there is matter present which generates a horizon with radius $R_h$, then we are saying that the semiclassical picture is recovered when we take the virtual fuzzballs to satisfy (\ref{bbtone}). 

\b

(b) As the universe expands, the horizon radius increases if $\rho_m$ is composed of dust or radiation. The virtual fuzzballs from different horizon  sized regions merge, so that $R_{max}$ increases with time. We assume that the merger process is rapid enough at this stage so that the virtual fuzzballs can be large enough to fill up the horizon. Thus at all steps in this process we have
\be
R_{max}=H^{-1}
\ee
and semiclassical physics will appear to be valid, by our earlier assumptions. Note that this evolution process is irreversible, as the larger fuzzballs formed by mergers occupy a bigger phase space than the unmerged fuzzballs. 

\b

(c) It is possible that at some point in this evolution the virtual fuzzballs are unable to merge rapidly enough to keep pace with the increasing size of the horizon. In this situation we will have
\be
R_{max}<H^{-1}
\label{bbttwo}
\ee
In a vacuum spacetime, if we had fuzzball sizes limited by $R_{max}$, then we argued that the spacetime developed a cosmological constant given by (\ref{bbel}); i.e, it developed a de Sitter curvature with length scale $\sim R_{max}$. In the present case with horizon $H^{-1}$, we conjecture that when $R_{max}$ falls below $H^{-1}$, then a cosmological constant $\Lambda$ will develop. Such a $\Lambda$ might explain the dark energy we see today. The condition may get triggered by inhomogeneities in $\rho_m$, which would give a resolution to the coincidence problem.  It has been argued that inhomogeneities in $\rho_m$ gives rise to an effective $\Lambda$, while other computations  have argued against this conclusion. Here we have a new ingredient: an inhomogeneity in the scale $R_{max}$ of virtual fuzzballs in the quantum gravity wavefunctional $\Psi$.  We can then ask  if such inhomogeneities can trigger the condition (\ref{bbttwo}) and thus a nonvanishing $\Lambda$ \cite{buchert,branmajum,waldrebut}. It would also be interesting if inhomogeneous values of the limit $R_{max}$ of virtual fuzzballs could act as dark matter. (For  reviews, see for example \cite{peebles,sami}. For observational constraints on the behavior of our cosmology today, see for example \cite{sola,anjan,amendola, staro}. For other approaches,   see for example \cite{mukhanov,dgp,vilenkin,kklt, kaloper,panda}.)

\section{Summary}\label{secsummary}

We have proposed a fundamental change in the way we should view the quantum gravity wavefunctional $\Psi$. Traditionally, we have expected that $\Psi$ has violent quantum fluctuations at the planck scale, but 
semiclassical physics at lower energies. This traditional picture leads to two problems: the information paradox for black holes and the cosmological constant problem.

Our proposal is that there is a new component in $\Psi$ that is crucial: the quantum fluctuations of extended compression-resistant structures; i.e, virtual fuzzballs. While individual fuzzballs of mass $M\gg m_p$ are heavy and so their virtual fluctuations are suppressed, the overall number of such fuzzballs is very large -- $Exp[S_{bek}(M)]$ -- and so the overall effect of this component of the wavefunction is important. 

In the case of the black hole, these virtual fluctuations generate a `pseuro-Rindler region' around the fuzzball depicted by the shaded area in fig.\ref{figrindlercausalityp}. The physics of this region can be modeled as a variable depth sea pictured in fig.\ref{fig3qq}. Low energy pulses (having a small height) do not notice the finite depth of the sea, but pulses of sufficiently large energy have a different behavior and evolve to fuzzballs, as depicted in fig.\ref{fig2qq}. Thus a horizon never forms, and we evade the information paradox. If we do not introduce the virtual fuzzball component of $\Psi$, then an infalling shell feels nothing special as it falls through its horizon, and it gets trapped inside this horizon (fig.\ref{fs}). Then we cannot resolve the information paradox without violating causality; i.e., without allowing signal propagation outside the light cone. 

Having arrived at the virtual fuzzball component of $\Psi$, we can now ask what role it plays in cosmology.\footnote{A model where actual on-shell fuzzballs give the matter in the universe was studied in \cite{masoumi}.} The quantum fluctuations of local quantum fields generates a large cosmological constant $\Lambda$, which we assume to be positive; by itself this $\Lambda$ would curve spacetime so that it has a very small curvature radius. Most traditional approaches to the cosmological constant problem have relied on some sort of fine tuning which cancels this large quantum contribution to $\Lambda$ against a bare $\Lambda$ of the opposite sign. We, however find a different mechanism for the emergence of spacetime with a large curvature radius. Suppose the virtual fuzzballs in $\Psi$ extend to a radius $R_{max}$. If the space were highly curved, then these fuzzballs would get squeezed to a smaller size. Since the fuzzballs are compression resistant, they do not allow this squeezing, and consequently flatten space so that the curvature radius becomes at least $R_{max}$ (fig.\ref{figc}). This proposal for cancelling the large apparent $\Lambda$ ties in well with the black hole problem, since we find that it allows us to bypass the `bags of gold' construction which is an alternative way of seeing the information paradox.

With this understanding of the virtual fuzzball component of $\Psi$, we conjectured a picture for the evolution of the wavefunction in cosmology. In the traditional semiclassical picture, an infinite dust universe can be regarded as the limit of a large homogeneous ball of dust sitting in asymptotically flat space. With the virtual fuzzball component of $\psi$, this limit is no longer smooth: the wavefunctional in the case of the infinite universe (C1) differs in structure from the wavefunctional in the infinite universe case (C2). In case C2, we have asymptotically flat spacetime, and thus near infinity we will have virtual fuzzballs with arbitrarily large mass $M$. If this dust ball is allowed to collapse (in order to describe a time reversed cosmology) then the dust ball will transition to fuzzball when it reaches its horizon radius. Any subsequent picture of local semiclassical physics can  then only be obtained through the conjecture of fuzzball complementarity. In case C1, we are forced to assume that the virtual fuzzballs only extend to a maximal radius $R_{max}$ which is less than or equal to the horizon radius. If $R_{max}$ equals the horizon radius, then we recover semiclassical physics, just as we recover flat spacetime semiclassical physics when we have fuzzballs of arbitrarily large radius $R$.  If $R_{max}$ is less than the horizon radius, then we will get a cosmological constant.

\section*{Acknowledgements}

I am grateful to Sumit Das, Bin Guo, Anupam Majumdar, Ali Masoumi, M. Sami, Anjan Sen, Joan Sola, Amitabh Virmani and Yogesh Srivastava for helpful discussions. This work is supported in part by a grant from the FQXi foundation.

 \newpage

\begin{appendix}

\refstepcounter{section}
\section*{\thesection \quad The causal horizon}

We assume that the universe is heading towards a big crunch, so that $\dot a(t)<0$. Consider the sphere around $\vec r=0$ with radius $r_1(t)$. The proper radius of this sphere is
\be
R_1(t)=a(t) r_1(t)
\ee
The area of the sphere is
\be
A=\Omega_{d-1} R_1^{d-1}=\Omega_{d-1} [a(t) r_1(t)]^{d-1}
\ee
Consider light rays travelling radially outwards from the sphere. Along such a ray
\be
a(t) dr=dt, ~~~{dr\over dt}={1\over a(t)}
\ee
Thus
\be
{dA\over dt}=\Omega_{d-1}(d-1)[a(t) r_1(t)]^{d-2}[\dot a (t) r_1(t)+a(t) \dot r_1(t)]
\ee
Thus the expansion of the outward directed null rays reaches zero when
\be
\dot a (t) r_1(t)+a(t) \dot r_1(t)=0
\ee
which gives
\be
{\dot a(t)\over a(t)}=-{\dot r_1(t)\over r_1(t)}=-{1\over a(t) r_1(t)}=-{1\over R_1(t)}
\ee
Thus we get a horizon when
\be
R_1=-\left ( {\dot a(t)\over a(t)}\right )^{-1}\equiv H^{-1}
\ee
where $H$ is the Hubble constant. We set
\be
R_h=H^{-1}
\label{bbtwentyq}
\ee
as the horizon that we will be interested in.

The energy density of the dust, measured in a local orthonormal frame where the dust is at rest,   has the form
 \be
 T^t{}_t=\rho_0={C\over (a(t))^d}
 \ee
 The cosmological metric has a the expansion
 \be
 a(t)=a_0 t^{2\over d}=\left ( {4\pi dGC\over (d-1)}\right )^{1\over d} t^{2\over d}
 \label{qten}
 \ee
At the boundary $R_b$ we wish to join our ball to the Schwarzschild metric. Our question is: what will be the mass parameter for this Schwarzschild metric? 
 The Schwarzschild metric has the form
 \be
 ds^2_S=-(1-\left ({R_0\over R}\right ) ^{d-2})dT^2+ {dR^2\over (1-\left ({R_0\over R}\right ) ^{d-1}} + R^2 d\Omega_{d-2}^2
 \ee
so our goal is to find $R_0$, as a function of the cosmological time $t$ when we wish to make the match.

To match the cosmological coordinates to the Schwarzschild coordinates, we note that
 \be
 R_b=a(t) r_b=a_0 t^{2\over d} r_b
 \label{qthree}
 \ee
 We can also equate proper times along the dust particles that are at the boundary of the ball. In the cosmological metric, these particles move along fixed $\vec r$, with
 \be
 d\tau=dt
 \ee
In the Schwarzschild metric, let the trajectory of the boundary be $R_b(T_b)$; here we assume that the surface of the ball is still outside the horizon radius.  Then
 \be
d\tau^2=\left  (1-\left ({R_0\over R_b}\right ) ^{d-2}\right )dT_b^2-{dR_b^2\over \left (1-\left ({R_0\over R_b}\right ) ^{d-2}\right )}
\label{qsix}
\ee
In the Schwarzschild metric, the quantity 
\be
E\equiv -g_{TT}(R_b) {dT_b\over d\tau}
\ee
is conserved along the infall trajectory. We have taken a flat cosmology, which corresponds to the particle falling in from infinity with zero initial velocity. At infinity we would therefore get
\be
E={dT_b\over d\tau}=1
\ee
Thus at all points along the infall
\be
{dT_b\over d\tau}=\left ( -g_{TT}(R_b)\right ) ^{-1} = \left (1-\left ({R_0\over R_b}\right ) ^{d-2}\right )^{-1}
\label{qfive}
\ee
From (\ref{qthree}) we have
\be
{dR_b\over d\tau}={\dot a(t) r_b }{dt\over d\tau}=\dot a(t) r_b
\label{qfour}
\ee
Substituting (\ref{qfive}) and (\ref{qfour}) in (\ref{qsix}) we get
\be
1=\left  (1-\left ({R_0\over R_b}\right ) ^{d-2}\right )^{-1}[1-(\dot a (t) r_b)^2]
\ee
This gives
\be
\dot a (t) r_b=\left ({R_0\over R_b}\right ) ^{d-2\over 2}
\ee
Substituting the value of $\dot a(t)$ from (\ref{qten}) we get
\be
R_0= (\left ( {16\pi GC\over d(d-1)}\right )r_b^d)^{{1\over (d-2)}}
\label{qtw}
\ee

In the cosmological setting it is not immediately obvious what we should call the mass $m$ inside the ball $R=R_b$. We have the rest mass of the dust particles, but we also have a negative gravitational potential energy from the gravitational attraction between the dust particles, and a positive kinetic energy from the motion of these particles. The case of a flat dust cosmology can be mapped to the interior of a dust ball collapsing from rest at infinity. For such a dust ball the potential and kinetic energies cancel, and the entire mass comes from just the rest energy of the dust particles. Thus we set
\be
m=\rho_0 {1\over d}\Omega_{d-1}(a(t)r_b)^{d}={C\over (a(t))^d} {1\over d}\Omega_{d-1}(a(t)r_b)^{d}={C} {1\over d}\Omega_{d-1} r_b^{d}
\label{qel}
\ee
We now check that this is the same as the mass $M$ that we would get for the exterior Schwarzschild solution if we match  the cosmological ball $R<R_b$ to asymptotically flat spacetime. From  (\ref{qtw}) and  (\ref{qel}) we find
\be
R_0^{d-2}={16\pi Gm\over (d-1)\Omega_{d-1}}
\label{qthir}
\ee
This is exactly the value of $R_0$ that we get for a Schwarzschild solution of mass
\be
M=m
\ee
in $d+1$ spacetime dimensions. In particular, for $d=3$ we get $R_0=2GM$.

\end{appendix}


\end{document}